\documentclass[american,aps,prb,reprint,longbibliography,superscriptaddress]{revtex4-2}
\usepackage[T1]{fontenc}
\usepackage[utf8]{inputenc}
\usepackage{babel}
\usepackage{verbatim}
\usepackage{booktabs}
\usepackage{amsmath}
\usepackage{amssymb}
\usepackage{graphicx}
\usepackage[
    pdfusetitle,
    bookmarks=false,
    breaklinks=true,
    pdfborder={0 0 0},
    pdfborderstyle={},
    backref=false,
    colorlinks=true]{hyperref}
\hypersetup{
    citecolor=purple,
    linkcolor=purple,
    urlcolor=black
}

\graphicspath{{fig/}}
\usepackage{todonotes}

\makeatletter

\providecommand{\tabularnewline}{\\}

\usepackage{orcidlink}

\usepackage{enumitem}
\usepackage{textcomp}

\ifdefined\LYXINTERNAL
\else
	\usepackage{newtxtext,newtxmath} % better Times + math matching PRB more or less
\fi

\def\fnum@figure{\textbf{Fig.~\thefigure}}
\def\fnum@table{\textbf{Tab.~\thetable}}

\makeatother

\begin{document}
\title{Exact Multi-Valley Envelope Function Theory of Valley Splitting in Si/SiGe Nanostructures}

\author{Lasse Ermoneit
\orcidlink{0009-0006-0329-0164}}
\email{ermoneit@wias-berlin.de}
\affiliation{Weierstrass Institute for Applied Analysis and Stochastics (WIAS),
Anton-Wilhelm-Amo-Str. 39,
10117 Berlin,
Germany}

\author{Abel Thayil
\orcidlink{0000-0002-3438-5901}}
\affiliation{Weierstrass Institute for Applied Analysis and Stochastics (WIAS),
Anton-Wilhelm-Amo-Str. 39,
10117 Berlin,
Germany}

\author{Thomas Koprucki
\orcidlink{0000-0001-6235-9412}}
\affiliation{Weierstrass Institute for Applied Analysis and Stochastics (WIAS),
Anton-Wilhelm-Amo-Str. 39,
10117 Berlin,
Germany}

\author{Markus Kantner
\orcidlink{0000-0003-4576-3135}}
\email{kantner@wias-berlin.de}
\affiliation{Weierstrass Institute for Applied Analysis and Stochastics (WIAS),
Anton-Wilhelm-Amo-Str. 39,
10117 Berlin,
Germany}

\begin{abstract}
Valley splitting in strained Si/SiGe quantum wells is a central parameter for silicon spin qubits and is commonly described with envelope-function and effective-mass theories.
These models provide a computationally efficient continuum description and have been shown to agree well with atomistic approaches when the confinement potential is slowly varying on the lattice scale.
In modern Si/SiGe heterostructures with atomically sharp interfaces and engineered Ge concentration profiles, however, the slowly varying potential approximation underlying conventional (local) envelope-function theory is challenged.
We formulate an exact multi-valley envelope-function model by combining Burt--Foreman-type envelope-function theory, which does not rely on the assumption of a slowly varying potential, with a valley-sector decomposition of the Brillouin zone.
This construction enforces band-limited envelopes, which satisfy a set of coupled integro-differential equations with a non-local potential energy operator.
Using degenerate perturbation theory, we derive the intervalley coupling matrix element within this non-local model and prove that it is strictly invariant under global shifts of the confinement potential (choice of reference energy).
We then show that the conventional local envelope model generically violates this invariance due to spectral leakage between valley sectors, leading to an unphysical energy-reference dependence of the intervalley coupling.
The resulting ambiguity is quantified by numerical simulations of various engineered Si/SiGe heterostructures.
Finally, we propose a simple spectrally filtered local approximation that restores the energy-reference invariance exactly and provides a good approximation to the exact non-local theory. 
\end{abstract}

\maketitle

\section{Introduction}

Silicon-based spin qubits in strained Si/SiGe quantum dots (QDs) \cite{Burkard2023,Zwanenburg2013} are a promising platform for scalable quantum processors, owing to their long coherence times \cite{Yoneda2017,Tyryshkin2011}, compatibility with industrial fabrication \cite{Koch2025,Neyens2024,George2025}, and potential for high-fidelity gate operations and readout \cite{Mills2022,Noiri2022,Xue2022}.
A key challenge for silicon qubits is the small and device-dependent valley splitting, \emph{i.e.}, the energy gap between the lowest conduction band valley states, which can lead to uncontrolled valley excitations, spin-valley mixing, and spin dephasing \cite{Yang2013,Zhang2020,Losert2024}.
Considerable effort has been made to enhance the valley splitting by engineering high-quality Si/SiGe interfaces with low alloy disorder \cite{DegliEsposti2024}.
Moreover, epitaxial growth of unconventional heterostructures including, \emph{e.g.}, sharp Ge spikes \cite{McJunkin2021} or oscillating Ge concentration profiles within the quantum well (QW) \cite{McJunkin2022,Feng2022,Woods2024,Gradwohl2025}, can provide deterministic enhancements of the valley splitting.

Envelope function theory (EFT) and effective-mass models provide a computationally efficient framework for describing electrons in semiconductor nanostructures, in which the wave function is written as a slowly varying envelope modulating Bloch states.
In regimes where the envelope description is applicable, these continuum-scale models have been shown to agree well with atomistic approaches such as empirical tight-binding \cite{Niquet2009,McJunkin2022,Losert2023} and density functional theory \cite{Cvitkovich2024,Cvitkovich2026}.
The assumption of a slowly varying potential, which is central to conventional EFT \cite{Kohn1955,Fritzsche1962,Baldereschi1970,Ning1971,Pantelides1974,Shindo1976,Debernardi2006,Hui2013,Pendo2013,Gamble2015}, becomes increasingly challenged in engineered heterostructures with sharp features.
To address this limitation, Burt and Foreman developed an exact EFT that does not require a slowly varying potential, leading to a system of coupled integro-differential equations with a non-local potential energy operator \cite{Burt1988,Burt1994,Foreman1995,Foreman1996}.
Klymenko \emph{et al.} extended this framework to multi-valley semiconductors by decomposing the Brillouin zone into valley-specific sectors, which has been applied to donor qubits in silicon \cite{Klymenko2014,Klymenko2015}.

In this work, we investigate valley splitting in Si/SiGe heterostructures using the exact multi-valley EFT and compare its mathematical properties and numerical predictions with those of the commonly used local (conventional) model.
We show that the local multi-valley envelope function model can violate invariance under global shifts of the confinement potential (choice of reference energy) when the envelope develops Fourier components outside the  corresponding valley-specific sector of the Brillouin zone.
This introduces an unphysical contribution to the intervalley coupling matrix element, making the predicted valley splitting dependent on the arbitrary choice of energy reference.
Physically, a constant potential offset should only shift intravalley energies and leave the intervalley coupling invariant.
We trace this inconsistency to the failure of the local model to strictly enforce the valley-sector (band-limited) Fourier support of the envelopes, and we prove that it is exactly resolved within the Burt--Foreman type approach used here.

The paper is organized as follows:
In Sec.~\ref{sec: Exact Multi-Valley Envelope Function Theory}, we review the derivation of the Burt--Foreman type envelope function model for multi-valley semiconductors.
In Sec.~\ref{sec: valley splitting}, we develop degenerate perturbation theory for the valley splitting within the non-local eigenvalue problem and analyze gauge invariance (\emph{i.e.}, invariance under global energy shifts) of the intervalley coupling in local and exact EFT.
Section~\ref{sec: numerical results} presents numerical results for electrons in gate-defined QDs for several Si/SiGe heterostructure designs and benchmarks a simple spectrally filtered local approximation against the exact non-local theory.
Technical aspects, including the effective one-dimensional QW reduction and proofs, are provided in the Appendix.

\begin{figure}[t]
\includegraphics[width=1\columnwidth]{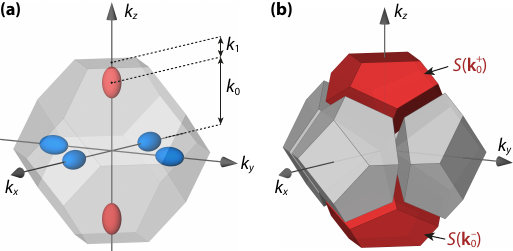}
\caption{\textbf{(a)}~First Brillouin zone of the face-centered cubic (fcc) lattice.
In biaxially strained SiGe/Si/SiGe QWs grown in {[}001{]}
direction, the degeneracy between the six equivalent conduction band minima near the $X$-points is lifted.
The two valley states at $\mathbf{k}_{0}^{\pm}=\left(0,0,\pm k_{0}\right)^{T}$ (shown in red) are energetically far below the other four higher-energy valley states (shown in blue).
\textbf{(b)}~Decomposition of the first Brillouin zone of the fcc lattice into non-overlapping valley-specific sectors $S\left(\mathbf{k}_{0}\right)$.
The sectors of the two low-energy valleys highlighted in red define the two-valley model (\ref{eq: two-valley model-2}).
}
\label{fig:grid-brillouin}
\end{figure}

\section{Exact Multi-Valley Envelope Function Theory \label{sec: Exact Multi-Valley Envelope Function Theory}}

This section derives the exact multi-valley EFT used throughout the paper.
By projecting the envelope functions onto non-overlapping, valley-specific sectors of the Brillouin zone, we obtain a unique decomposition of the microscopic wave function and arrive at a coupled set of envelope equations in which the confinement potential acts as a non-local integral operator.
We then derive the corresponding single-band effective-mass model and recover the conventional local EFT in the limit where the valley-sector projection is dropped.

\subsection{Microscopic Schrödinger Equation}

The electronic state of a single electron in a semiconductor nanostructure
is described by the stationary microscopic Schrödinger equation
\begin{equation}
\left(-\frac{\hbar^{2}}{2m_{0}}\nabla^{2}+V\left(\mathbf{r}\right)+U\left(\mathbf{r}\right)\right)\Psi_{\alpha}\left(\mathbf{r}\right)=E_{\alpha}\Psi_{\alpha}\left(\mathbf{r}\right),\label{eq:micro-se}
\end{equation}
where $m_{0}$ is the vacuum electron mass, $V$ is the lattice-periodic crystal potential, and $U$ is a non-periodic mesoscopic confinement potential.
The index $\alpha$ labels the
energy eigenvalues $E_{\alpha}$ and wave functions $\Psi_{\alpha}$.
Throughout this paper, we assume that the potential $V$ describes a perfectly periodic Si crystal (diamond structure), whereas the effects of Ge atoms forming the heterostructure are phenomenologically included in the mesoscopic potential $U$ \cite{Thayil2025}.
The potential $U$ also includes the potential of the gate-induced electric fields defining the QD.

\subsection{Multi-Valley Envelope Wave Function}

We expand the microscopic wave function in Bloch eigenstates of the lattice-periodic potential $V$ and derive an exact equation for the corresponding envelope functions.
Throughout this work, \emph{exact} refers to the formally exact envelope-function representation (without slowly varying potential approximation) and not to atomistic correctness of the chosen potential $U$.
This goes beyond conventional EFT and effective-mass theory \cite{Kohn1955, Fritzsche1962, Ning1971, Pantelides1974, Shindo1976, Hui2013}, which relies on the assumption that the mesoscopic confinement potential varies slowly on the scale of the lattice.
This assumption is challenged by sharp interfaces~\cite{DegliEsposti2024}, narrow Ge spikes~\cite{McJunkin2021}, and highly oscillatory ``wiggle-well'' profiles~\cite{McJunkin2022, Gradwohl2025} considered in modern Si/SiGe qubit heterostructures. 
To avoid this limitation, we employ the exact EFT developed by Burt and Foreman \cite{Burt1988,Burt1994,Foreman1995,Foreman1996}, which yields a set of coupled integro-differential equations with a non-local potential energy operator.
Klymenko \emph{et al.} \cite{Klymenko2014,Klymenko2015} extended this framework to multi-valley semiconductors by decomposing the Brillouin zone into non-overlapping, valley-specific sectors, as illustrated in Fig.~\ref{fig:grid-brillouin}.

We consider the ansatz 
\begin{equation}
\Psi_{\alpha}\left(\mathbf{r}\right)=\sum_{n,\mathbf{k}_{0}}u_{n,\mathbf{k}_{0}}\left(\mathbf{r}\right)F_{n,\mathbf{k}_{0},\alpha}\left(\mathbf{r}\right),\label{eq:decomposition}
\end{equation}
where $u_{n,\mathbf{k}_{0}}$ are the lattice-periodic Bloch factors and $F_{n,\mathbf{k}_{0},\alpha}$ are envelope wave functions labeled by the band index $n$ and valley index $\mathbf{k}_{0}$.
A detailed justification of the ansatz is given in Appendix~\ref{sec: ansatz justification}. 
The Bloch factors satisfy
\begin{equation}
\left(\frac{\hbar^{2}}{2m_{0}}\left(-\mathrm{i}\nabla+\mathbf{k}\right)^{2}+V\left(\mathbf{r}\right)\right)u_{n,\mathbf{k}}\left(\mathbf{r}\right)=E_{n,\mathbf{k}}u_{n,\mathbf{k}}\left(\mathbf{r}\right)\label{eq:bloch-eq}
\end{equation}
evaluated at the valley wave vectors $\mathbf{k}=\mathbf{k}_{0}$.
We consider biaxially strained Si, for which the two valleys at $\pm\mathbf{k}_{0}=\left(0,0,\pm k_{0}\right)^{T}$ are energetically
separated from the remaining four conduction band valleys \cite{VandeWalle1986,Schaeffler1997}.
The strained lattice-periodic potential can be modeled, \emph{e.g.}, using empirical pseudopotentials \cite{Rieger1993,Ungersboeck2007b,Thayil2025}. 
The lattice-periodic Bloch factors can be expressed as a Fourier series over reciprocal lattice vectors $\mathbf{G}$
\begin{equation}
u_{n,\mathbf{k}_{0}}\left(\mathbf{r}\right)=\sum_{\mathbf{G}}\mathrm{e}^{\mathrm{i}\mathbf{G}\cdot\mathbf{r}}c_{n,\mathbf{k}_{0}}\left(\mathbf{G}\right).\label{eq: Bloch Fourier series}
\end{equation}
To ensure uniqueness of the ansatz (\ref{eq:decomposition}), the
Fourier expansion of the envelope functions $F_{n,\mathbf{k}_{0},\alpha}$
is restricted to plane wave components from their respective valley-sectors $S\left(\mathbf{k}_0\right)$ as 
\begin{align}
F_{n,\mathbf{k}_{0},\alpha}\left(\mathbf{r}\right) & =\sum_{\mathbf{K}\in S\left(\mathbf{k}_{0}\right)}\mathrm{e}^{\mathrm{i}\mathbf{K}\cdot\mathbf{r}}F_{n,\mathbf{k}_{0},\alpha}\left(\mathbf{K}\right)\label{eq: envelope F}\\
 & =\sum_{\mathbf{K}\in S\left(\mathbf{k}_{0}\right)}\mathrm{e}^{\mathrm{i}\mathbf{K}\cdot\mathbf{r}}\sum_{\mathbf{G}}c_{n,\mathbf{k}_{0}}^{*}\left(\mathbf{G}\right)\Psi_{\alpha}\left(\mathbf{K}+\mathbf{G}\right),\nonumber 
\end{align}
where ${\mathbf{K}=\left(\mathbf{k}+\mathbf{k}_{0}\right)\in S\left(\mathbf{k}_{0}\right)}$.
%confined to momentum space components belonging to the valley-specific sector $S(\vb{k}_0)$ of the first Brillouin zone, thereby preventing overcompleteness.
\begin{comment}
Here we introduced the indicator function $\chi_{S(\vb{k}_{0})}$
in momentum space with regard to the sector $S(\vb{k}_{0})$. 
\begin{align}
\chi_{S(\vb{k}_{0})}(\vb{K}) & =\begin{cases}
1, & \text{if }\vb{K}\in S(\vb{k}_{0}),\\
0, & \text{else.}
\end{cases}
\end{align}
\end{comment}
The dominant component of the envelope wave function $F_{n,\mathbf{k}_{0},\alpha}$
is a plane wave with wave vector $\mathbf{k}_{0}$.
We define a \emph{slowly varying envelope}
\begin{equation}
f_{n,\mathbf{k}_{0},\alpha}\left(\mathbf{r}\right)=\mathrm{e}^{-\mathrm{i}\mathbf{k}_{0}\cdot\mathbf{r}}F_{n,\mathbf{k}_{0},\alpha}\left(\mathbf{r}\right),
\label{eq: slowly varying envelope f}
\end{equation}
where the rapid valley-scale oscillations are absorbed.

\subsection{Orthonormality and Completeness}

The eigenstates of the full single-electron problem (\ref{eq:micro-se})
form an orthonormal basis set that is complete with respect to the
$L^{2}$ norm over the crystal volume $V_{c}$ \begin{subequations}\label{eq:psi-orthonomal-complete}
\begin{align}
\int_{V_{c}}\mathrm{d}^{3}r\,\Psi_{\alpha}^{*}\left(\mathbf{r}\right)\Psi_{\alpha'}\left(\mathbf{r}\right) & =\delta_{\alpha,\alpha'},\label{eq:psi-orthonormal}\\
\sum_{\alpha}\Psi_{\alpha}\left(\mathbf{r}\right)\Psi_{\alpha}^{*}\left(\mathbf{r}'\right) & =\delta\left(\mathbf{r}-\mathbf{r}'\right).\label{eq:psi-complete}
\end{align}
\end{subequations}For the Bloch factors with a fixed wave vector
$\mathbf{k}$, we assume orthonormalization over the primitive unit
cell volume $\Omega_{p}$ and completeness on the space of lattice-periodic
functions \begin{subequations}\label{eq:u-orthonomal-complete}
\begin{align}
\frac{1}{\Omega_{p}}\int_{\Omega_{p}}\mathrm{d}^{3}r\,u_{n,\mathbf{k}}^{*}\left(\mathbf{r}\right)u_{n',\mathbf{k}}\left(\mathbf{r}\right)
&=
\delta_{n,n'},\label{eq:u-orthonormal}\\
\frac{1}{\Omega_{p}}
\sum_{n}u_{n,\mathbf{k}}^{*}\left(\mathbf{r}\right)u_{n,\mathbf{k}}\left(\mathbf{r}'\right) & =\delta\left(\mathbf{r}-\mathbf{r}'\right).\label{eq:u-complete}
\end{align}
\end{subequations}The orthogonality relations (\ref{eq:psi-orthonormal})
and (\ref{eq:u-orthonormal}) imply orthonormality of the envelopes
after summation over all bands and valleys \begin{subequations}\label{eq:F-orthonomal-complete}
\begin{equation}
\sum_{n,\mathbf{k}_{0}}\int_{V_{c}}\mathrm{d}^{3}r\,F_{n,\mathbf{k}_{0},\alpha}^{*}\left(\mathbf{r}\right)F_{n,\mathbf{k}_{0},\alpha'}\left(\mathbf{r}\right)=\delta_{\alpha,\alpha'}.\label{eq: F-orthonormal}
\end{equation}
Moreover, the envelopes are complete on the space of band-limited
functions (restricted to Fourier components from the valley-sectors $S\left(\mathbf{k}_{0}\right)$) 
\begin{equation}
\sum_{\alpha}F_{n,\mathbf{k}_{0},\alpha}\left(\mathbf{r}\right)F_{n',\mathbf{k}_{0}',\alpha}^{*}\left(\mathbf{r}'\right)=\delta_{n,n^{\prime}}\delta_{\mathbf{k}_{0},\mathbf{k}_{0}'}\Delta_{\mathbf{k}_{0}}\left(\mathbf{r}-\mathbf{r}'\right).\label{eq: F-complete}
\end{equation}
\end{subequations}
In Eq.~(\ref{eq: F-complete}), we have introduced a truncated (sector-projected) delta function adapted to the valley-sector $S\left(\mathbf{k}_{0}\right)$ \cite{Klymenko2015}
\begin{equation}
\Delta_{\mathbf{k}_{0}}\left(\mathbf{r}-\mathbf{r}'\right)=\frac{1}{V_{c}}\sum_{\mathbf{K}\in S\left(\mathbf{k}_{0}\right)}\mathrm{e}^{\mathrm{i}\mathbf{K}\cdot\left(\mathbf{r}-\mathbf{r}'\right)},\label{eq: Delta low pass}
\end{equation}
where the summation range in momentum space is restricted to $\mathbf{K}\in S\left(\mathbf{k}_{0}\right)$.
Hence, $\Delta_{\mathbf{k}_{0}}$ is the kernel of the projector onto $S\left(\mathbf{k}_0\right)$, admitting only plane wave components from this valley-sector of the Brillouin zone.
The function plays a key role in the exact multi-valley EFT considered here, especially to guarantee gauge invariance (under shifts of the reference energy) of the intervalley coupling as pointed out in Sec.~\ref{sec: Gauge Invariance} below.
For the envelope wave functions $F_{n,\mathbf{k}_{0},\alpha}$ compatible with Eq.~(\ref{eq: envelope F}), the valley-matched function $\Delta_{\mathbf{k}_{0}}\left(\mathbf{r}-\mathbf{r}'\right)$ acts as a conventional Dirac delta function
\begin{equation}
\int_{V_{c}}\mathrm{d}^{3}r'\,\Delta_{\mathbf{k}_{0}}\left(\mathbf{r}-\mathbf{r}'\right)F_{n,\mathbf{k}_{0}',\alpha}\left(\mathbf{r}'\right)=\delta_{\mathbf{k}_{0},\mathbf{k}_{0}'}F_{n,\mathbf{k}_{0},\alpha}\left(\mathbf{r}\right).\label{eq: Delta function convolution property}
\end{equation}
When extending the summation range to the full momentum space (dropping the restriction to valley-sectors), the truncated delta function approaches the conventional Dirac delta function.
We note the properties
\begin{equation}
\Delta_{\mathbf{k}_{0}}^{*}
\left(\mathbf{r}-\mathbf{r}'\right)
=
\Delta_{\mathbf{k}_{0}}
\left(\mathbf{r}'-\mathbf{r}\right)
=
\Delta_{-\mathbf{k}_{0}}
\left(\mathbf{r}-\mathbf{r}'\right).
\label{eq: Delta low pass properties}
\end{equation}

\subsection{Multi-Valley Envelope Equation}

Substituting the ansatz (\ref{eq:decomposition}) into the microscopic Schrödinger equation (\ref{eq:micro-se}) yields, without invoking a slowly varying potential approximation, 
%, after some algebra
%\cite{Thayil2025}, 
an eigenvalue equation for the envelopes \cite{Thayil2025}:
\begin{align}
E_{\alpha}F_{n,\mathbf{k}_{0},\alpha}\left(\mathbf{r}\right) & =-\frac{\hbar^{2}}{2m_{0}}\nabla^{2}F_{n,\mathbf{k}_{0},\alpha}\left(\mathbf{r}\right)\label{eq: envelope equation F}\\
 & -\mathrm{i}\sum_{n'}\frac{\hbar}{m_{0}}\mathbf{p}_{n,n'}\left(\mathbf{k}_{0}\right)\cdot\nabla F_{n',\mathbf{k}_{0},\alpha}\left(\mathbf{r}\right)\nonumber \\
 & +\sum_{n'}\frac{\hbar^{2}}{2m_{0}}\left(T_{n,n'}\left(\mathbf{k}_{0}\right)+V_{n,n'}\left(\mathbf{k}_{0}\right)\right)F_{n',\mathbf{k}_{0},\alpha}\left(\mathbf{r}\right)\nonumber \\
 & +\sum_{n',\mathbf{k}_{0}'}\int_{V_{c}}\mathrm{d}^{3}r'\,U_{\mathbf{k}_{0},\mathbf{k}_{0}'}^{n,n'}\left(\mathbf{r},\mathbf{r}'\right)F_{n',\mathbf{k}_{0}',\alpha}\left(\mathbf{r}'\right). \nonumber
\end{align}
The matrix elements of the momentum, kinetic energy, and the lattice-periodic part of the potential energy read
\begin{align*}
\mathbf{p}_{n,n'}\left(\mathbf{k}_{0}\right) & =\sum_{\mathbf{G}}c_{n,\mathbf{k}_{0}}^{*}\left(\mathbf{G}\right)\hbar\mathbf{G}c_{n',\mathbf{k}_{0}}\left(\mathbf{G}\right),\\
T_{n,n'}\left(\mathbf{k}_{0}\right) & =\sum_{\mathbf{G}}c_{n,\mathbf{k}_{0}}^{*}\left(\mathbf{G}\right)\frac{\hbar^{2}G^{2}}{2m_{0}}c_{n',\mathbf{k}_{0}}\left(\mathbf{G}\right),\\
V_{n,n'}\left(\mathbf{k}_{0}\right) & =\sum_{\mathbf{G},\mathbf{G}'}c_{n,\mathbf{k}_{0}}^{*}\left(\mathbf{G}\right)V\left(\mathbf{G}-\mathbf{G}'\right)c_{n',\mathbf{k}_{0}}\left(\mathbf{G}'\right).
\end{align*}
The non-local term in Eq.~(\ref{eq: envelope equation F}) involves the matrix elements of the mesoscopic potential
\begin{align}
U_{\mathbf{k}_{0},\mathbf{k}_{0}'}^{n,n'}\left(\mathbf{r},\mathbf{r}'\right) & =\int_{V_{c}}\mathrm{d}^{3}r''\,\Delta_{\mathbf{k}_{0}}\left(\mathbf{r}-\mathbf{r}''\right)u_{n,\mathbf{k}_{0}}^{*}\left(\mathbf{r}''\right)U\left(\mathbf{r}''\right)\nonumber \\
 & \hphantom{=\int_{V_{c}}\mathrm{d}^{3}r'\,}\times u_{n',\mathbf{k}_{0}'}\left(\mathbf{r}''\right)\Delta_{\mathbf{k}_{0}'}\left(\mathbf{r}''-\mathbf{r}'\right).\label{eq: mesoscopic potential matrix element U}
\end{align}
The matrix element $U_{\mathbf{k}_{0},\mathbf{k}_{0}'}^{n,n'}$ has a project-multiply-project structure and gives rise to a coupling between valley states.

Rewriting the envelope equation (\ref{eq: envelope equation F}) in
terms of the slowly varying envelopes $f_{n,\mathbf{k}_{0},\alpha}$,
we obtain using the Bloch equation~(\ref{eq:bloch-eq}) at $\mathbf{k}=\mathbf{k}_{0}$
the system 
\begin{align}
E_{\alpha} & f_{n,\mathbf{k}_{0},\alpha}\left(\mathbf{r}\right)=\hat{H}_{n,\mathbf{k}_{0}}\left(\mathbf{r}\right)f_{n,\mathbf{k}_{0},\alpha}\left(\mathbf{r}\right)\label{eq:env-eq1}\\
 & =\left(-\frac{\hbar^{2}}{2m_{0}}\nabla^{2}+E_{n,\mathbf{k}_{0}}\right)f_{n,\mathbf{k}_{0},\alpha}\left(\mathbf{r}\right)\nonumber \\
 & -\mathrm{i}\frac{\hbar}{m_{0}}\sum_{n'}\left(\mathbf{p}_{n,n'}\left(\mathbf{k}_{0}\right)+\hbar\mathbf{k}_{0}\delta_{n,n'}\right)\cdot\nabla f_{n',\mathbf{k}_{0},\alpha}\left(\mathbf{r}\right)\nonumber \\
 & +\sum_{n',\mathbf{k}_{0}'}\int_{V_{c}}\mathrm{d}^{3}r'\,u_{\mathbf{k}_{0},\mathbf{k}_{0}'}^{n,n'}\left(\mathbf{r},\mathbf{r}'\right)f_{n',\mathbf{k}_{0}',\alpha}\left(\mathbf{r}'\right)\nonumber 
\end{align}
where we introduced the transformed non-local kernel
\begin{equation}
u_{\mathbf{k}_{0},\mathbf{k}_{0}'}^{n,n'}\left(\mathbf{r},\mathbf{r}'\right)=\mathrm{e}^{-\mathrm{i}\left(\mathbf{k}_{0}\cdot\mathbf{r}-\mathbf{k}_{0}'\cdot\mathbf{r}'\right)}U_{\mathbf{k}_{0},\mathbf{k}_{0}'}^{n,n'}\left(\mathbf{r},\mathbf{r}'\right).\label{eq: mesoscopic potential matrix element u}
\end{equation}
The kernel has the symmetry properties
\begin{equation}
\left(u_{\mathbf{k}_{0},\mathbf{k}_{0}'}^{n,n'}\left(\mathbf{r},\mathbf{r}'\right)\right)^{*}=u_{\mathbf{k}_{0}',\mathbf{k}_{0}}^{n',n}\left(\mathbf{r}',\mathbf{r}\right)=u_{-\mathbf{k}_{0},-\mathbf{k}_{0}'}^{n,n'}\left(\mathbf{r},\mathbf{r}'\right),\label{eq: properties u}
\end{equation}
which follows from Eq.~(\ref{eq: Delta low pass properties}) and
the identity for the lattice-periodic Bloch factors $u_{n,-\mathbf{k}_{0}}\left(\mathbf{r}\right)=u_{n,\mathbf{k}_{0}}^{*}\left(\mathbf{r}\right)$.
In combination with suitable boundary
conditions (\emph{i.e.}, homogeneous Dirichlet or periodic
boundary conditions), the symmetry~(\ref{eq: properties u}) along with completeness~(\ref{eq: F-complete}) guarantees self-adjointness of the Hamiltonian $\hat{H}_{n,\mathbf{k}_{0}}$, \emph{i.e.},
\begin{align}
\sum_{n,\mathbf{k}_{0}}\int_{V_{c}} & \mathrm{d}^{3}r\,f_{n,\mathbf{k}_{0},\alpha'}^{*}\left(\mathbf{r}\right)\left(\hat{H}_{n,\mathbf{k}_{0}}f_{n,\mathbf{k}_{0},\alpha}\left(\mathbf{r}\right)\right)=\\
 & =\sum_{n,\mathbf{k}_{0}}\int_{V_{c}}\mathrm{d}^{3}r\,\left(\hat{H}_{n,\mathbf{k}_{0}}f_{n,\mathbf{k}_{0},\alpha'}\left(\mathbf{r}\right)\right)^{*}f_{n,\mathbf{k}_{0},\alpha}\left(\mathbf{r}\right)\nonumber 
\end{align}
with $\hat{H}_{n,\mathbf{k}_{0}}=\hat{H}_{n,\mathbf{k}_{0}}^{\dagger}$.
This ensures a real-valued energy spectrum $E_{\alpha}\in\mathbb{R}$.

\subsection{Single-Band Model (Effective Mass Approximation)}

The system of envelope equations (\ref{eq:env-eq1}) is reduced to an effective single-band model by perturbatively decoupling the conduction band ${n=c}$ from remote bands labeled by ${r\neq c}$, \emph{e.g.}, via a Schrieffer--Wolff transformation or Löwdin renormalization.
Following \cite{Thayil2025}, we obtain
\begin{align}
E_{\alpha} & f_{c,\mathbf{k}_{0},\alpha}\left(\mathbf{r}\right)=\label{eq: single-band-model}\\
 & =-\frac{\hbar^{2}}{2}\nabla\cdot\left(m_{c,\mathbf{k}_{0}}^{-1}\nabla f_{c,\mathbf{k}_{0},\alpha}\left(\mathbf{r}\right)\right)+E_{c,\mathbf{k}_{0}}f_{c,\mathbf{k}_{0},\alpha}\left(\mathbf{r}\right)\nonumber \\
 & \phantom{=}+\sum_{\mathbf{k}_{0}^{\prime}}\int_{V_{c}}\mathrm{d}^{3}r'\,u_{\mathbf{k}_{0},\mathbf{k}_{0}'}^{c,c}\left(\mathbf{r},\mathbf{r}'\right)f_{c,\mathbf{k}_{0}',\alpha}\left(\mathbf{r}'\right)\nonumber 
\end{align}
with the inverse effective mass tensor of the conduction band valley
states given as
\begin{equation}
m_{c,\mathbf{k}_{0}}^{-1}=\frac{1}{m_{0}}I_{3\times3}+\frac{2}{m_{0}^{2}}\sum_{r}\frac{\mathbf{p}_{c,r}\left(\mathbf{k}_{0}\right)\otimes\mathbf{p}_{r,c}^{\dagger}\left(\mathbf{k}_{0}\right)}{E_{c,\mathbf{k}_{0}}-E_{r,\mathbf{k}_{0}}}.\label{eq: effective mass}
\end{equation}
We emphasize that only the interband coupling has been eliminated
in Eq.~(\ref{eq: single-band-model}), whereas the intervalley coupling
structure mediated by the mesoscopic potential was kept intact.

\subsection{Local Limit: Conventional Envelope Function Theory}

The conventional envelope function model is recovered by dropping the valley-sector projection.
This is equivalent to replacing the truncated delta functions~(\ref{eq: Delta low pass}) in the mesoscopic potential matrix elements (\ref{eq: mesoscopic potential matrix element u}) by conventional Dirac delta functions 
\begin{equation*}
\Delta_{\mathbf{k}_{0}}\left(\mathbf{r}-\mathbf{r}'\right)\to\delta\left(\mathbf{r}-\mathbf{r}'\right).
\end{equation*}
As a consequence, the uniqueness of the decomposition (\ref{eq:decomposition}) is lost, which matters whenever the envelopes $f_{n,\mathbf{k}_{0},\alpha}$ develop short-wavelength components outside of their valley sectors.
In this \emph{local limit}, the matrix elements (\ref{eq: mesoscopic potential matrix element u}) reduce to 
\begin{align*}
\left.u_{\mathbf{k}_{0},\mathbf{k}_{0}'}\left(\mathbf{r},\mathbf{r}'\right)\right\vert _{\mathrm{loc}} & =\mathrm{e}^{-\mathrm{i}\left(\mathbf{k}_{0}-\mathbf{k}_{0}'\right)\cdot\mathbf{r}}u_{\mathbf{k}_{0}}^{*}\left(\mathbf{r}\right)U\left(\mathbf{r}\right)u_{\mathbf{k}_{0}'}\left(\mathbf{r}\right)\times\\
 & \hphantom{={}}\times\delta\left(\mathbf{r}-\mathbf{r}'\right).
\end{align*}
Approximating the rapidly oscillating Bloch factors in the intravalley matrix elements ($\mathbf{k}_{0}=\mathbf{k}_{0}'$) as 
\begin{equation*}
\left|u_{\mathbf{k}_{0}}\left(\mathbf{r}\right)\right|^{2}\approx1,
\end{equation*}
one arrives at the local envelope equation model 
\begin{equation}
E_{\alpha}f_{c,\mathbf{k}_{0},\alpha}^{\mathrm{loc}}\left(\mathbf{r}\right)=\left(\hat{H}_{0}^{\mathrm{loc}}+\hat{H}_{1}^{\mathrm{loc}}\right)f_{c,\mathbf{k}_{0},\alpha}^{\mathrm{loc}}\left(\mathbf{r}\right)\label{eq: local envelope equation}
\end{equation}
with
\begin{align*}
\hat{H}_{0}^{\mathrm{loc}} & =-\frac{\hbar^{2}}{2}\nabla\cdot m_{c,\mathbf{k}_{0}}^{-1}\nabla+E_{c,\mathbf{k}_{0}}+U\left(\mathbf{r}\right),\\
\hat{H}_{1}^{\mathrm{loc}} & =\sum_{\mathbf{k}_{0}'\neq\mathbf{k}_{0}}\mathrm{e}^{-\mathrm{i}\left(\mathbf{k}_{0}-\mathbf{k}_{0}'\right)\cdot\mathbf{r}}u_{\mathbf{k}_{0}}^{*}\left(\mathbf{r}\right)U\left(\mathbf{r}\right)u_{\mathbf{k}_{0}'}\left(\mathbf{r}\right).
\end{align*}
The model (\ref{eq: local envelope equation}) has been widely employed in the silicon qubit literature \cite{Feng2022,Saraiva2009,Woods2024,Thayil2025,Friesen2007}.
In this model, the envelopes are typically dominated by long-wavelength components, but there is no projection mechanism that enforces valley-sector band limitation.
In the following section, we point out that it is this inconsistency of local EFT, which leads to a violation of gauge invariance and ambiguity in the magnitude of the intervalley coupling.

\section{Theory of Valley Splitting \label{sec: valley splitting}}

\subsection{Coupled Two-Valley Envelope Model}

We consider a biaxially strained Si/SiGe QW grown along the {[}001{]} crystallographic direction and restrict the model to the two low-energy conduction band valleys at $\mathbf{k}_{0}^{\pm}=\left(0,0,\pm k_{0}\right)^{T}$, see Fig.~\ref{fig:grid-brillouin}\,(b).
Omitting the band index, we collect the corresponding slowly varying envelopes in
\begin{equation*}    
\mathbf{f}_{\alpha}\left(\mathbf{r}\right)=\left(\begin{array}{c}
f_{\mathbf{k}_{0}^{+},\alpha}\left(\mathbf{r}\right)\\
f_{\mathbf{k}_{0}^{-},\alpha}\left(\mathbf{r}\right)
\end{array}\right).
\end{equation*}
We assume identical conduction band edges $E_{c}=E_{c,\mathbf{k}_{0}^{\pm}}$ and effective mass tensors $m_{c}=m_{c,\mathbf{k}_{0}^{\pm}}$ for the two valleys.
The resulting two-component eigenvalue problem reads
\begin{equation}
E_{\alpha}\mathbf{f}_{\alpha}\left(\mathbf{r}\right)=\hat{H}\mathbf{f}_{\alpha}\left(\mathbf{r}\right),
\label{eq: two-valley model-2}
\end{equation}
where the non-local Hamiltonian is decomposed into an intravalley and an intervalley contribution:
\begin{equation*}
\hat{H}=\hat{H}_{0}+\hat{H}_{1}.
\end{equation*}
The intravalley Hamiltonian reads
\begin{align}
\hat{H}_{0}\mathbf{f}_{\alpha}\left(\mathbf{r}\right) & =\left(-\frac{\hbar^{2}}{2}\nabla\cdot\left(m_{c}^{-1}\nabla\right)+E_{c}\right)\mathbf{f}_{\alpha}\left(\mathbf{r}\right)\label{eq: intravalley H0}\\
 & \phantom{=}+\int_{V_{c}}\mathrm{d}^{3}r'\,\left(\begin{array}{cc}
u_{\mathbf{k}_{0}^{+},\mathbf{k}_{0}^{+}}\left(\mathbf{r},\mathbf{r}'\right) & 0\\
0 & u_{\mathbf{k}_{0}^{-},\mathbf{k}_{0}^{-}}\left(\mathbf{r},\mathbf{r}'\right)
\end{array}\right)\mathbf{f}_{\alpha}\left(\mathbf{r}'\right)\nonumber \\
 & =\left(\begin{array}{cc}
\hat{H}_{0}^{+} & 0\\
0 & \hat{H}_{0}^{-}
\end{array}\right)
\mathbf{f}_{\alpha}\left(\mathbf{r}\right),
\nonumber 
\end{align}
where $\hat{H}_{0}^{\pm}$ denote the corresponding non-local single-valley operators acting on the envelopes.
The intervalley Hamiltonian reads
\begin{equation}
\hat{H}_{1}\mathbf{f}_{\alpha}\left(\mathbf{r}\right)=\int_{V_{c}}\mathrm{d}^{3}r'\,\left(\begin{array}{cc}
0 & u_{\mathbf{k}_{0}^{+},\mathbf{k}_{0}^{-}}\left(\mathbf{r},\mathbf{r}'\right)\\
u_{\mathbf{k}_{0}^{-},\mathbf{k}_{0}^{+}}\left(\mathbf{r},\mathbf{r}'\right) & 0
\end{array}\right)\mathbf{f}_{\alpha}\left(\mathbf{r}'\right),\label{eq: intervalley H1}
\end{equation}
where hermiticity is ensured by $u_{\mathbf{k}_{0}^{+},\mathbf{k}_{0}^{-}}^{*}\left(\mathbf{r},\mathbf{r}'\right)=u_{\mathbf{k}_{0}^{-},\mathbf{k}_{0}^{+}}\left(\mathbf{r},\mathbf{r}'\right)$.
We take $m_{c}=\mathrm{diag}\left(m_{t},m_{t},m_{l}\right)$,
with transverse mass $m_{t}$ and longitudinal mass $m_{l}$.

Because the mesoscopic potential enters through valley-projected non-local kernels, the problem (\ref{eq: two-valley model-2}) is a non-local generalization of the widely used two-valley effective-mass model \cite{Friesen2007, Thayil2025, Woods2024, Saraiva2009, Feng2022}.
In contrast to conventional (local) EFT, however, the two intravalley operators $\hat{H}_{0}^{\pm}$ are not identical due to the different mesoscopic potential matrix elements $u_{\mathbf{k}_{0}^{+},\mathbf{k}_{0}^{+}}$ and $u_{\mathbf{k}_{0}^{-},\mathbf{k}_{0}^{-}}$.
Instead, the two components are related by complex conjugation
\begin{align}
\left(\hat{H}_{0}^{+}\right)^{*} & =\hat{H}_{0}^{-},\label{eq: H0 components complex conj}
\end{align}
which follows from $u_{\mathbf{k}_{0}^{-},\mathbf{k}_{0}^{-}}^{*}\left(\mathbf{r},\mathbf{r}'\right)=u_{\mathbf{k}_{0}^{+},\mathbf{k}_{0}^{+}}\left(\mathbf{r},\mathbf{r}'\right)$,
see Eq.~(\ref{eq: properties u}).

\subsection{Invariance under Reference Energy Shifts \label{sec: Gauge Invariance}}

The physics of the coupled two-valley system (\ref{eq: two-valley model-2}) must be invariant under a constant offset of the confinement potential (choice of reference energy) 
by an arbitrary constant  $U_{0}$ 
\begin{align*}
U\left(\mathbf{r}\right) & \to U\left(\mathbf{r}\right)+U_{0},\\
E_{\alpha} & \to E_{\alpha}+U_{0}.
\end{align*}
In particular, the intervalley coupling must remain invariant, even though $\hat{H}_{1}$ depends explicitly on $U$.

In the non-local model (\ref{eq: two-valley model-2}), the intravalley
($\mathbf{k}_{0}=\mathbf{k}_{0}'$) and the intervalley ($\mathbf{k}_{0}\neq\mathbf{k}_{0}'$)
matrix elements of the mesoscopic potential $u_{\mathbf{k}_{0},\mathbf{k}_{0}'}\left(\mathbf{r},\mathbf{r}'\right)$
transform differently under the global shift.
By a straightforward computation, we obtain 
\begin{equation}
u_{\mathbf{k}_{0},\mathbf{k}_{0}'}\left(\mathbf{r},\mathbf{r}'\right)\xrightarrow{U\to U+U_{0}}u_{\mathbf{k}_{0},\mathbf{k}_{0}'}\left(\mathbf{r},\mathbf{r}'\right)+U_{0}\delta_{\mathbf{k}_{0},\mathbf{k}_{0}'}\Delta_{\mathbf{k}_{0}}\left(\mathbf{r}-\mathbf{r}'\right),\label{eq: gauge transform u}
\end{equation}
see  Appendix \ref{sec: proof gauge invariance} for details.
Consequently, the valley-sector projection ensures that the offset $U_0$ shifts only the diagonal (intravalley) terms, while it cancels exactly in the off-diagonal (intervalley) terms:
\begin{align*}
\hat{H}_{0} & \xrightarrow{U\to U+U_{0}}\hat{H}_{0}+U_{0}\,I_{2\times2},\\
\hat{H}_{1} & \xrightarrow{U\to U+U_{0}}\hat{H}_{1}.
\end{align*}
The shift in $\hat{H}_{0}$ is compensated by the simultaneous shift of the energy eigenvalues $E_{\alpha}\to E_{\alpha}+U_{0}$,
leaving the eigenvalue problem (\ref{eq: two-valley model-2}) invariant.
The same argument extends directly to the multi-band system (\ref{eq:env-eq1}).

Crucially, this invariance relies on the valley-sector projection built into the exact EFT.
By contrast, the conventional local envelope equation (\ref{eq: local envelope equation}) omits this projection, and the transformation $U\to U+U_0$ produces an additional contribution to the intervalley Hamiltonian,
\begin{align*}
\hat{H}_{1}^{\mathrm{loc}} & \xrightarrow{U\to U+U_{0}}\hat{H}_{1}^{\mathrm{loc}}+U_{0}\sum_{\mathbf{k}_{0}'\neq\mathbf{k}_{0}}\mathrm{e}^{-\mathrm{i}\left(\mathbf{k}_{0}-\mathbf{k}_{0}'\right)\cdot\mathbf{r}}u_{\mathbf{k}_{0}}^{*}\left(\mathbf{r}\right)u_{\mathbf{k}_{0}'}\left(\mathbf{r}\right),
\end{align*}
\emph{i.e.}, an unphysical dependence of the intervalley coupling on the choice of energy reference.
In Sec.~\ref{sec: numerical results} we quantify the resulting ambiguity for several heterostructures using the perturbative expression for the intervalley coupling matrix element derived next.

\subsection{Perturbation Theory \label{sec: perturbation theory}}

%We develop the perturbation theory of the valley splitting within the exact (non-local) envelope equation theory.
We treat the intervalley coupling Hamiltonian $\hat{H}_{1}$ as a weak perturbation and write
\begin{align*}
E_{\alpha}\mathbf{f}_{\alpha} & \left(\mathbf{r}\right)=\left(\hat{H}_{0}+\varepsilon\hat{H}_{1}\right)\mathbf{f}_{\alpha}\left(\mathbf{r}\right)
\end{align*}
with a formal small parameter $\varepsilon$.
Expanding the envelope and the energy eigenvalue to first order
\begin{align*}
\mathbf{f}_{\alpha}\left(\mathbf{r}\right)
&=
\mathbf{f}_{\alpha}^{\left(0\right)}\left(\mathbf{r}\right)+\varepsilon\mathbf{f}_{\alpha}^{\left(1\right)}\left(\mathbf{r}\right)+O\left(\varepsilon^{2}\right),\\
E_{\alpha}
&=
E_{\alpha}^{\left(0\right)}+\varepsilon E_{\alpha}^{\left(1\right)}+O\left(\varepsilon^{2}\right)
\end{align*}
yields the standard degenerate-perturbation-theory problem.
We assume orthonormalization of the full envelopes
\begin{align*}
\int_{V_{c}}\mathrm{d}^{3}r\,\mathbf{f}_{\alpha}^{\dagger}\left(\mathbf{r}\right)\cdot\mathbf{f}_{\alpha'}\left(\mathbf{r}\right) & =\delta_{\alpha,\alpha'}
\end{align*}
and of their zeroth-order components
\begin{align*}
\int_{V_{c}}\mathrm{d}^{3}r\,\left(\mathbf{f}_{\alpha}^{\left(0\right)}\left(\mathbf{r}\right)\right)^{\dagger}\cdot\mathbf{f}_{\alpha'}^{\left(0\right)}\left(\mathbf{r}\right) & =\delta_{\alpha,\alpha'}.
\end{align*}

At $\varepsilon=0$, the two valleys are decoupled
\begin{equation}
E_{\alpha}^{\left(0\right)}\mathbf{f}_{\alpha}^{\left(0\right)}\left(\mathbf{r}\right)=\hat{H}_{0}\mathbf{f}_{\alpha}^{\left(0\right)}\left(\mathbf{r}\right).\label{eq: perturbation theory f - eps 0}
\end{equation}
Using the two linearly independent basis vectors
\begin{align*}
\mathbf{f}_{+,\alpha}^{\left(0\right)}\left(\mathbf{r}\right) & =\left(\begin{array}{c}
f_{\mathbf{k}_{0}^{+},\alpha}^{\left(0\right)}\left(\mathbf{r}\right)\\
0
\end{array}\right), & \mathbf{f}_{-,\alpha}^{\left(0\right)}\left(\mathbf{r}\right) & =\left(\begin{array}{c}
0\\
f_{\mathbf{k}_{0}^{-},\alpha}^{\left(0\right)}\left(\mathbf{r}\right)
\end{array}\right),
\end{align*}
the system is reduced to two scalar eigenproblems
\begin{subequations}
\begin{align}
\hat{H}_{0}^{+}f_{\mathbf{k}_{0}^{+},\alpha}^{\left(0\right)}\left(\mathbf{r}\right) & =E_{\alpha}^{\left(0\right)}f_{\mathbf{k}_{0}^{+},\alpha}^{\left(0\right)}\left(\mathbf{r}\right),\label{eq: zeroth order H_0^+}\\
\hat{H}_{0}^{-}f_{\mathbf{k}_{0}^{-},\alpha}^{\left(0\right)}\left(\mathbf{r}\right) & =E_{\alpha}^{\left(0\right)}f_{\mathbf{k}_{0}^{-},\alpha}^{\left(0\right)}\left(\mathbf{r}\right).\label{eq: zeroth order H_0^-}
\end{align}
\end{subequations}
Because $\hat{H}_{0}^{-}=\left(\hat{H}_{0}^{+}\right)^*$, the spectrum is two-fold degenerate and the eigenfunctions may be chosen such that
\begin{align*}
\left(f_{\mathbf{k}_{0}^{+},\alpha}^{\left(0\right)}\left(\mathbf{r}\right)\right)^{*} & =f_{\mathbf{k}_{0}^{-},\alpha}^{\left(0\right)}\left(\mathbf{r}\right)
\end{align*}
with normalization
\begin{equation*}
\int_{V_{c}}\mathrm{d}^{3}r\,\left|f_{\mathbf{k}_{0}^{+},\alpha}^{\left(0\right)}\left(\mathbf{r}\right)\right|^{2}=\int_{V_{c}}\mathrm{d}^{3}r\,f_{\mathbf{k}_{0}^{+},\alpha}^{\left(0\right)}\left(\mathbf{r}\right)f_{\mathbf{k}_{0}^{-},\alpha}^{\left(0\right)}\left(\mathbf{r}\right)=1.
\end{equation*}

The first-order problem in $O\left(\varepsilon\right)$ reads 
\begin{equation}
E_{\alpha}^{\left(0\right)}\mathbf{f}_{\alpha}^{\left(1\right)}\left(\mathbf{r}\right)+E_{\alpha}^{\left(1\right)}\mathbf{f}_{\alpha}^{\left(0\right)}\left(\mathbf{r}\right)=\hat{H}_{0}\mathbf{f}_{\alpha}^{\left(1\right)}\left(\mathbf{r}\right)+\hat{H}_{1}\mathbf{f}_{\alpha}^{\left(0\right)}\left(\mathbf{r}\right).\label{eq: perturbation theory f - eps 1}
\end{equation}
Since the spectrum of the unperturbed single-valley problem is two-fold degenerate, we employ degenerate perturbation theory to compute the first-order correction of the energy eigenvalues. Thus we consider a superposition of the orthogonal modes spanning the degenerate subspace
$\mathbf{f}_{\alpha}^{\left(0\right)}\left(\mathbf{r}\right)=\eta_{+}\mathbf{f}_{+,\alpha}^{\left(0\right)}\left(\mathbf{r}\right)+\eta_{-}\mathbf{f}_{-,\alpha}^{\left(0\right)}\left(\mathbf{r}\right)$
and derive a system for the amplitudes $\eta_{\pm}$.
After projection onto the basis modes, we arrive at
\begin{equation*}
\left(\begin{array}{cc}
0 & \Delta\\
\Delta^{*} & 0
\end{array}\right)\left(\begin{array}{c}
\eta_{+}\\
\eta_{-}
\end{array}\right)=E_{\alpha}^{\left(1\right)}\left(\begin{array}{c}
\eta_{+}\\
\eta_{-}
\end{array}\right)
\end{equation*}
with the complex-valued intervalley coupling matrix element
\begin{equation}
\Delta=\int_{V_{c}}\mathrm{d}^{3}r\,\int_{V_{c}}\mathrm{d}^{3}r'\,f_{\mathbf{k}_{0}^{+},\alpha}^{\left(0\right)*}\left(\mathbf{r}\right)u_{\mathbf{k}_{0}^{+},\mathbf{k}_{0}^{-}}\left(\mathbf{r},\mathbf{r}'\right)f_{\mathbf{k}_{0}^{-},\alpha}^{\left(0\right)}\left(\mathbf{r}'\right).\label{eq: Delta (with f)}
\end{equation}
The  first-order eigenvalues are $E_{\alpha}^{\left(1\right)}=\pm\left|\Delta\right|$,
leading to the valley splitting
\begin{equation*}
E_{\mathrm{VS}}=2\left|\Delta\right|.
\end{equation*}
In Eq.~(\ref{eq: Delta (with f)}), the intervalley coupling matrix element is expressed as a two-fold integral over a non-local kernel.
Using the definitions of the mesoscopic potential matrix elements, this expression can be rewritten in a form that resembles the conventional EFT result, but with the crucial difference that the envelopes are restricted to their valley-specific Brillouin zone sectors.
Substituting the expressions (\ref{eq: mesoscopic potential matrix element U})
and (\ref{eq: mesoscopic potential matrix element u}) for the matrix
element, we arrive at
\begin{align*}
\Delta & =\int_{V_{c}}\mathrm{d}^{3}r\,\int_{V_{c}}\mathrm{d}^{3}r'\,f_{\mathbf{k}_{0}^{+},\alpha}^{\left(0\right)*}\left(\mathbf{r}\right)\mathrm{e}^{-\mathrm{i}\left(\mathbf{k}_{0}^{+}\cdot\mathbf{r}-\mathbf{k}_{0}^{-}\cdot\mathbf{r}'\right)}\times\\
 & \hphantom{=\int_{V_{c}}\mathrm{d}^{3}r\,\int_{V_{c}}\mathrm{d}^{3}r'\,}\times U_{\mathbf{k}_{0}^{+},\mathbf{k}_{0}^{-}}\left(\mathbf{r},\mathbf{r}'\right)f_{\mathbf{k}_{0}^{-},\alpha}^{\left(0\right)}\left(\mathbf{r}'\right)\\
 & =\int_{V_{c}}\mathrm{d}^{3}r\,\int_{V_{c}}\mathrm{d}^{3}r'\,\int_{V_{c}}\mathrm{d}^{3}r''\,\left(F_{\mathbf{k}_{0}^{+},\alpha}^{\left(0\right)}\left(\mathbf{r}\right)\right)^{*}\Delta_{\mathbf{k}_{0}^{+}}\left(\mathbf{r}-\mathbf{r}''\right)\times\\
 & \hphantom{=\{\}}\times u_{\mathbf{k}_{0}^{+}}^{*}\left(\mathbf{r}''\right)U\left(\mathbf{r}''\right)u_{\mathbf{k}_{0}^{-}}\left(\mathbf{r}''\right)\Delta_{\mathbf{k}_{0}^{-}}\left(\mathbf{r}''-\mathbf{r}'\right)F_{\mathbf{k}_{0}^{-},\alpha}^{\left(0\right)}\left(\mathbf{r}'\right)
\end{align*}
with $F_{\mathbf{k}_{0}^{\pm},\alpha}^{\left(0\right)}\left(\mathbf{r}\right) = \mathrm{e}^{\mathrm{i}\mathbf{k}_{0}^{\pm}\cdot\mathbf{r}}f_{\mathbf{k}_{0}^{\pm},\alpha}^{\left(0\right)}\left(\mathbf{r}\right)$.
Using the convolution property (\ref{eq: Delta function convolution property}), we obtain 
\begin{equation}
\Delta=\int_{V_{c}}\mathrm{d}^{3}r\,\mathrm{e}^{-2\mathrm{i}\mathbf{k}_{0}\cdot\mathbf{r}}f_{\mathbf{k}_{0}^{+},\alpha}^{\left(0\right)*}\left(\mathbf{r}\right)u_{\mathbf{k}_{0}^{+}}^{*}\left(\mathbf{r}\right)U\left(\mathbf{r}\right)u_{\mathbf{k}_{0}^{-}}\left(\mathbf{r}\right)f_{\mathbf{k}_{0}^{-},\alpha}^{\left(0\right)}\left(\mathbf{r}\right),\label{eq: Delta (with f - simplified)}
\end{equation}
which is formally identical to the expression obtained within conventional EFT \cite{Thayil2025}.
The distinction is that in the exact EFT the envelopes are valley-sector projected, so that
\begin{equation*}
\Delta \xrightarrow{U\to U+U_{0}} \Delta.
\end{equation*}
By contrast, local EFT evaluates the same expression using envelopes that are not constrained to a single valley sector.
This sector leakage permits an unphysical dependence of $\Delta$ on global potential offsets and therefore violates invariance with respect to the choice of the reference energy.

\begin{table}
\begin{tabular*}{1\columnwidth}{@{\extracolsep{\fill}}lll}
\toprule 
\textbf{Symbol} & \textbf{Description} & \textbf{Value}\tabularnewline
\midrule
$a_{0}$ & Si lattice constant & $0.543\,\mathrm{nm}$\tabularnewline
$\Delta E_{c}$ & Si/Ge conduction band offset & $0.5\,\mathrm{eV}$\tabularnewline
$\Omega_{a}$ & atomic volume & $\left(a_{0}/2\right)^{3}$\tabularnewline
$k_{0}$ & valley wave number & $0.8394\times2\pi/a_{0}$\tabularnewline
$m_{t}$ & transverse effective mass & $0.209\,m_{0}$\tabularnewline
$m_{l}$ & longitudinal effective mass & $0.909\,m_{0}$\tabularnewline
$h$ & thickness of the QW domain & $75\,\mathrm{ML}$\tabularnewline
$\sigma_{u}$, $\sigma_{l}$ & upper and lower interface width & $0.5\,\mathrm{nm}$\tabularnewline
$X_{b}$ & nominal Ge concentration in barrier & $0.3$\tabularnewline
$\hbar\omega_{x}$, $\hbar\omega_{y}$ & circular QD orbital energy splitting & $3\,\mathrm{meV}$\tabularnewline
$F$ & vertical electric field & $3\,\mathrm{mV/nm}$\tabularnewline
\bottomrule
\end{tabular*}

\caption{Parameters used in the numerical simulations if not stated otherwise.
A complete list of parameters including also the empirical pseudopotential
model is given in Ref.~\cite{Thayil2025}. Here, $m_{0}$ is the
vacuum electron mass and $\mathrm{ML}=a_{0}/4$ is the silicon monolayer
thickness.}

\label{tab:parameters}
\end{table}

\section{Numerical Simulations
\label{sec: numerical results}}

In this section, we compute the valley splitting for several Si/SiGe heterostructure designs, including rapidly varying Ge concentration profiles relevant for electron qubits.
The computations use an effective one-dimensional model for the vertical confinement and extract the valley splitting from the perturbative intervalley coupling matrix element derived in Sec.~\ref{sec: perturbation theory}.
Throughout, we compare the exact non-local EFT with the conventional local approximation and with a projected-local (spectrally filtered) model introduced below.

\subsection{Mesoscopic Confinement Potential}

We consider a separable mesoscopic confinement potential of the form 
\begin{equation*}
U\left(\mathbf{r}\right) =U_{\perp}\left(z\right)+U_{\text{QD}}\left(x,y\right)
\end{equation*}
where $U_{\perp}\left(z\right)=U_{\text{het}}\left(z\right)+U_{F}\left(z\right)$
describes the vertical confinement due to the epitaxial layer stack and an applied electric field.
We model the heterostructure contribution as $U_{\mathrm{het}}=\Delta E_{c}\,X\left(z\right)$
where $\Delta E_{c}$ is the Si/Ge conduction band offset and $X\left(z\right)$ is the vertical Ge concentration
profile.
The electric-field potential is $U_{F}\left(z\right)=-e_{0}Fz$
with elementary charge $e_{0}$ and field strength $F$.
We write the concentration profile as
\begin{equation*}
X\left(z\right) = X_{\mathrm{QW}}\left(z\right)+X_{\mathrm{mod}}\left(z\right),
\end{equation*}
where $X_{\mathrm{QW}}\left(z\right)=X_{b}\,\left(1-\Xi\left(z\right)\right)$ represents a smoothed QW with barrier concentration $X_{b}$.
The QW shape function is
\begin{equation}
\Xi\left(z\right) =\frac{1}{2}\left(\tanh\left(\frac{z+h}{\sigma_{l}}\right)-\tanh\left(\frac{z}{\sigma_{u}}\right)\right), \label{eq: QW indicator}
\end{equation}
where $h$ is the QW thickness and $\sigma_{u,l}$ is the width of
the upper and lower interface, respectively.
The term $X_{\mathrm{mod}}\left(z\right)$ encodes engineered modifications of the QW profile, such as wiggle wells and Ge spikes considered below.

The in-plane confinement is modeled by a two-dimensional harmonic oscillator
\begin{equation*}
U_{\text{QD}}\left(x,y\right) =\frac{1}{2}m_{t}\omega_{x}^{2}x^{2}+\frac{1}{2}m_{t}\omega_{y}^{2}y^{2},
\end{equation*}
where $\omega_{x,y}$ controls the lateral QD size.
Effects of alloy disorder studied in Refs.~\cite{Thayil2025, Thayil2025c} are neglected in this work.

\begin{figure}
\includegraphics[width=1\columnwidth]{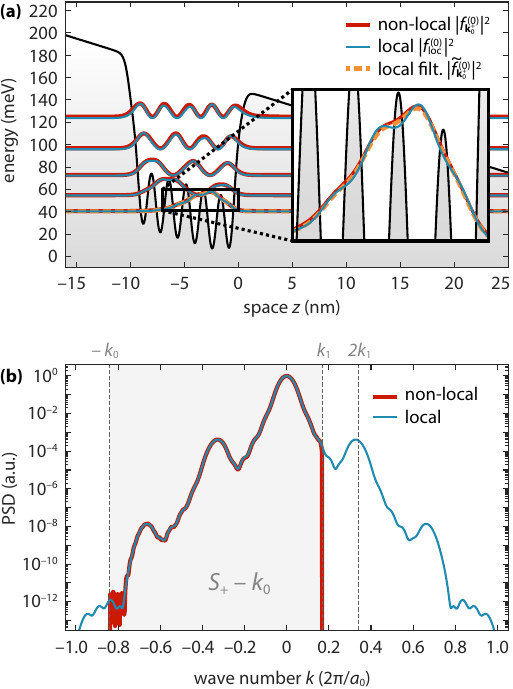}
\caption{\textbf{(a)}~Comparison of the eigenstates of a wiggle well heterostructure (\ref{eq: wiggle well}) with wave number $q=2k_{1}$ and Ge concentration
$X_{\mathrm{ww}}=0.1$ for a QW with thickness $h=72\,\mathrm{ML}$.
Absolute squares of slowly varying envelopes are shown as solutions of the the non-local eigenvalue problem (\ref{eq: effective Schroedinger equation Fourier space})
and its local approximation (\ref{eq: effective Schroedinger equation (local)}).
In addition, the projected local envelope according to Eq.~(\ref{eq: filtered envelope})
is shown as a dashed line for the ground state.
The inset shows a zoom on the vertical electron density distribution, which reveals a spurious modulation with frequency around $2k_{1}$ on the local
envelope, which is suppressed in the exact non-local model.
\textbf{(b)}~Power spectral densities (PSD) of the ground state envelopes.
The local envelope equation (\ref{eq: effective Schroedinger equation (local)})
does not impose any band-limitation on the envelope, such that it in general contains Fourier components outside of is valley-sector.
In contrast, the valley sector restriction is handled exactly in the non-local model (\ref{eq: effective Schroedinger equation Fourier space}).
Note that while $F_{\mathbf{k}_{0}^{+}}^{\left(0\right)}$
is restricted to $k\in\left(0,2\pi\left(1-\varepsilon_{z,z}\right)/a_{0}\right)$,
the domain is shifted by $-k_{0}$ to $k\in\left(-k_{0},+k_{1}\right)$
for the slowly varying envelope $f_{\mathbf{k}_{0}^{+}}^{\left(0\right)}$.
The PSDs are scaled to unity at $k=0$.
\label{fig: wave function}}
\end{figure}

\subsection{Effective One-Dimensional Schrödinger Equation}

The zeroth-order (single-valley) envelope $F_{\mathbf{k}_{0},\alpha}^{\left(0\right)}$ obeys an effective one-dimensional Schrödinger equation derived in Appendix~\ref{sec: effective one-dim Schroedinger}.
In Fourier space, the exact non-local equation reads
\begin{align}
E_{\alpha}F_{\mathbf{k}_{0}^{+},\alpha}^{\left(0\right)}\left(K\right) & =\left(\frac{\hbar^{2}K^{2}}{2m_{l}}+E_{c}+\frac{1}{2}\left(\hbar\omega_{x}+\hbar\omega_{y}\right)\right)F_{\mathbf{k}_{0}^{+},\alpha}^{\left(0\right)}\left(K\right)\nonumber \\
 &+\sum_{K'\in S_{+}}\sum_{n\in\mathbb{Z}}B_{n}U_{\perp}\left(K-K'+nG_{0,z}\right)F_{\mathbf{k}_{0}^{+},\alpha}^{\left(0\right)}\left(K'\right)\label{eq: effective Schroedinger equation Fourier space}
\end{align}
for $K\in S_{+}$, where $S_{+}$ denotes the valley-specific Brillouin zone segment.
The problem is formulated in Fourier space to enforce the restriction $K\in S_{+}$ explicitly.
See Appendix~\ref{sec: numerical method} for details on the numerical method.
The envelope of the opposite valley state follows from the identity
\begin{equation*}
F_{\mathbf{k}_{0}^{-},\alpha}^{\left(0\right)}\left(K\right) =
\left(F_{\mathbf{k}_{0}^{+},\alpha}^{\left(0\right)}\left(-K\right)\right)^{*}
\end{equation*}
for $K\in S_{-}$.
The corresponding slowly varying envelopes are obtained via a shift of the spectral support as 
${f_{\mathbf{k}_{0}^{\pm},\alpha}^{\left(0\right)}\left(K\right)
=
F_{\mathbf{k}_{0}^{\pm},\alpha}^{\left(0\right)}\left(K\pm k_0\right)}$
for ${K\in S_{\pm}}$.
Using the same approximation~(\ref{eq: Kronecker selection rule})
as in Appendix \ref{sec: effective one-dim Schroedinger}, the intervalley coupling matrix element is obtained as
\begin{equation}
\Delta =\sum_{n\in\mathbb{Z}}C_{n}^{\left(2\right)}\int\mathrm{d}z\,\mathrm{e}^{-\mathrm{i}\left(2k_{0}+nG_{0,z}\right)z}f_{\mathbf{k}_{0}^{+},\alpha}^{\left(0\right)*}\left(z\right)U_{\perp}\left(z\right)f_{\mathbf{k}_{0}^{-},\alpha}^{\left(0\right)}\left(z\right)\label{eq: Delta}
\end{equation}
with coefficients \cite{Thayil2025}
\begin{equation}
C_{n}^{\left(2\right)} =\sum_{\mathbf{G},\mathbf{G}'}c_{\mathbf{k}_{0}^{+}}^{*}\left(\mathbf{G}\right)c_{\mathbf{k}_{0}^{-}}\left(\mathbf{G}\right)\delta_{\mathbf{G}-\mathbf{G}',n\mathbf{G}_{0}}.\label{eq: C2 coefficient}
\end{equation}
Numerical values for $C_{n}^{\left(2\right)}$ including their shear strain dependence are given in Tab.~\ref{tab:Bloch coefficients}.

The results of the non-local model are compared with the local approximation, where the corresponding Schrödinger equation describing the uncoupled single-valley envelope $f_{\mathrm{loc},\alpha}^{\left(0\right)}\left(z\right)$
\begin{align}
E_{\alpha}f_{\mathrm{loc},\alpha}^{\left(0\right)}\left(z\right) & =-\frac{\hbar^{2}}{2m_{l}}\frac{\partial^{2}}{\partial z^{2}}f_{\mathrm{loc},\alpha}^{\left(0\right)}\left(z\right)\label{eq: effective Schroedinger equation (local)}\\
 & +\left(E_{c}+\frac{1}{2}\left(\hbar\omega_{x}+\hbar\omega_{y}\right)+U_{\perp}\left(z\right)\right)f_{\mathrm{loc},\alpha}^{\left(0\right)}\left(z\right)\nonumber 
\end{align}
is solved directly in position space.
Since the Hamiltonian in Eq.~(\ref{eq: effective Schroedinger equation (local)}) is real, the envelope may be chosen real and is identical for both valleys.
As a consequence, we can omit the valley index $\mathbf{k}_{0}^{\pm}$ for the slowly varying envelope
in the local approximation.
Unlike the exact non-local problem, the local model does not enforce valley-sector band limitation, so the envelope can acquire short-wavelength Fourier components outside the relevant Brillouin zone segment.

\subsection{Gauge Ambiguity in the Local Model}
The corresponding local approximation to the intervalley coupling matrix element $\Delta_{\mathrm{loc}}$
is obtained from (\ref{eq: Delta}) by substituting $f_{\mathbf{k}_{0},\alpha}^{\left(0\right)}\left(z\right)\to f_{\mathrm{loc},\alpha}^{\left(0\right)}\left(z\right)$
as
\begin{align}
\Delta_{\mathrm{loc}} & =\sum_{n}C_{n}^{\left(2\right)}\int\mathrm{d}z\,\mathrm{e}^{-\mathrm{i}\left(2k_{0}+nG_{0,z}\right)z}U_{\perp}\left(z\right)\left(f_{\mathrm{loc},\alpha}^{\left(0\right)}\left(z\right)\right)^{2}.\label{eq: Delta loc}
\end{align}
Because the local model is not invariant under global potential offsets, see Sec.~\ref{sec: Gauge Invariance}, $\Delta_{\mathrm{loc}}$ is not uniquely defined with respect to the choice of energy reference.
Under such an offset, the local intervalley coupling transforms as
\begin{equation*}
\Delta_{\mathrm{loc}}\xrightarrow{U\to U+U_{0}}\Delta_{\mathrm{loc}}+U_{0}R,
\end{equation*}
where the sensitivity to a constant shift is quantified by the ambiguity measure
\begin{equation}
R =\sum_{n}C_{n}^{\left(2\right)}\int\mathrm{d}z\,\mathrm{e}^{-\mathrm{i}\left(2k_{0}+nG_{0,z}\right)z}\left(f_{\mathrm{loc},\alpha}^{\left(0\right)}\left(z\right)\right)^{2}.
\label{eq: violation R}
\end{equation}
The predicted valley splitting
\begin{equation*}
E_{\mathrm{VS}}^{\mathrm{loc}}
=
2\left|\Delta_{\mathrm{loc}}\right|
\xrightarrow{U\to U+U_{0}}
2\left|\Delta_{\mathrm{loc}} + U_{0} R\right|
\leq
2\left|\Delta_{\mathrm{loc}}\right|+2U_{0}\left|R\right|
\end{equation*}
thus becomes energy-reference dependent.
Even if the local envelope is close to the non-local envelope with not much spectral weight outside of the admitted Brillouin zone segment, see Fig.~\ref{fig: wave function},
the effect on $E_{\mathrm{VS}}$ can be strong already for moderate $U_{0}$ of a few $100\,\mathrm{meV}$.

A practical consequence is that a local-EFT calculation can be made to ``agree'' with a measured valley splitting by an unphysical adjustment of the reference energy whenever $R$ is non-negligible.
Such offsets can arise in practice from different conventions for band-edge reference energies and potential offsets used across heterostructure and electrostatic modeling workflows, even though the underlying physics is unchanged.
This undermines the use of the local model for quantitative prediction and for validating experimental interpretations in sharply varying confinement landscapes.

\begin{figure}[t!]
\includegraphics[width=1\columnwidth]{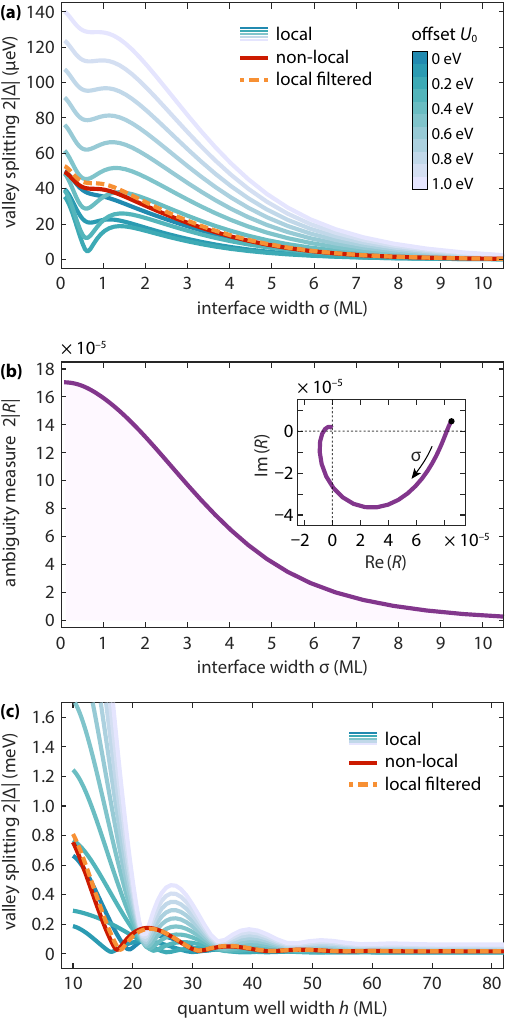}
\caption{Valley splitting in a conventional QW with smoothed interfaces.
\textbf{(a)}~Dependence of the valley splitting on the QW interface width $\sigma$.
The local model shows a strong (unphysical) dependence on the reference energy, which is illustrated by global offsets ranging from $U_{0}=0\,\mathrm{eV}$ to $U_{0}=1\,\mathrm{eV}$ (color-coded).
The projected local envelope approach (dashed, orange) closely approximates the exact result (red, solid).
\textbf{(b)}~The magnitude of the ambiguity measure $2\left|R\right|$ increases with decreasing interface width.
The inset shows a parametric plot of $R$ in the complex plane.
\textbf{(c)}~Valley splitting as a function of the QW width $h$.
The projected-local model remains in very good agreement with the non-local result, while the local model deviates strongly for thin wells where the slowly varying potential approximation breaks down.
\label{fig: conventional QW}}
\end{figure}

\subsection{Projected Local Model via Spectral Filtration}
%mod
To approximately restore valley-sector band limitation within the local model, we project the local envelope onto the corresponding valley sector,
\begin{equation}
\tilde{f}_{\mathbf{k}_{0},\alpha}^{\left(0\right)}\left(z\right)  =\int\mathrm{d}z'\,\Delta_{k_{0}}\left(z-z'\right)\mathrm{e}^{-ik_{0}\left(z-z'\right)}f_{\mathrm{loc},\alpha}^{\left(0\right)}\left(z'\right)\label{eq: filtered envelope}
\end{equation}
for $\mathbf{k}_{0}\in\left\{ \mathbf{k}_{0}^{+},\mathbf{k}_{0}^{-}\right\}$
followed by normalization.

The resulting projected-local approximation to the intervalley coupling matrix element is
\begin{align}
\Delta_{\mathrm{loc}}^{\mathrm{filtered}} & =\sum_{n\in\mathbb{Z}}C_{n}^{\left(2\right)}\int\mathrm{d}z\,\mathrm{e}^{-\mathrm{i}\left(2k_{0}+nG_{0,z}\right)z}\times\label{eq: Delta-loc filtered}\\
 & \hphantom{=\sum_{n\in\mathbb{Z}}C_{n}^{\left(2\right)}\int\mathrm{d}z\,}\times\tilde{f}_{\mathbf{k}_{0}^{+},\alpha}^{\left(0\right)*}\left(z\right)U_{\perp}\left(z\right)\tilde{f}_{\mathbf{k}_{0}^{-},\alpha}^{\left(0\right)}\left(z\right). \nonumber 
\end{align}
By construction, $\Delta_{\mathrm{loc}}^{\mathrm{filtered}}$ is invariant under global potential offsets, but it may deviate quantitatively from the exact non-local result. Below we benchmark this approximation for several heterostructure designs.

\subsection{Simulation Results
\label{sec: Simulation Results}}

\subsubsection{Interface Width}

We first consider a conventional QW with $X\left(z\right)=X_{\mathrm{QW}}\left(z\right)$
in a vertical electric field and vary the width $\sigma$ of the Si/SiGe interface (here we assume $\sigma=\sigma_{u}=\sigma_{l}$).
Figure~\ref{fig: conventional QW}\,(a) compares the valley splitting obtained from the exact non-local model, the conventional local approximation, and the projected-local (filtered) model.
For smooth interfaces, the ambiguity measure $\left|R\right|$ is small, see Fig.~\ref{fig: conventional QW}\,(b),
and all approaches yield similar results.
In contrast, for sharp interfaces on the scale of a few monolayers the local model develops a pronounced energy-reference dependence, leading to large deviations from the exact non-local prediction.
Projecting the local envelope onto the valley sector via Eq.~(\ref{eq: filtered envelope}) largely removes this inconsistency and yields close agreement with the exact result across the range of $\sigma$ considered.

\subsubsection{Quantum Well Width}
Next, we compute the valley splitting of a conventional QW in an electric field as a function of the well width $h$, see Fig.~\ref{fig: conventional QW}\,(c).
For thin QWs, the confinement potential varies rapidly and the slowly varying approximation underlying the local model breaks down \cite{Foreman1996}.
Correspondingly, the local prediction becomes strongly energy-reference dependent and deviates from the exact non-local result, while the projected-local model remains in good agreement.
The plot reveals several minima of the valley splitting at specific well widths, where the sinc-like Fourier spectrum of the finite-width QW potential has very small amplitudes at wave numbers governing the intervalley coupling (in particular near $2 k_0$ and $2 k_1$).

\subsubsection{Electric Field Dependency}

\begin{figure}
\includegraphics[width=1\columnwidth]{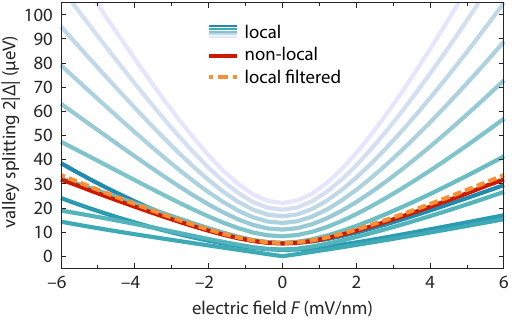}
\caption{Valley splitting as a function of the vertical electric field $F$ for a mirror-symmetric QW.
The exact non-local and projected-local models satisfy $E_{\mathrm{VS}}\left(F\right)=E_{\mathrm{VS}}\left(-F\right)$ as required by symmetry.
The conventional local model exhibits an unphysical asymmetry and a strong dependence on the reference energy (illustrated by offsets $U_{0}$ with the same color-coding as in Fig.~\ref{fig: conventional QW}).
\label{fig: electric field dependency}}
\end{figure}

Figure~\ref{fig: electric field dependency} shows the valley splitting as a function of the vertical electric field $F$ for a mirror-symmetric Ge concentration profile ($\sigma_u=\sigma_l$).
For a spatially symmetric heterostructure, the results must satisfy $E_{\mathrm{VS}}\left(F\right)=E_{\mathrm{VS}}\left(-F\right)$.
In the simulations the QW is placed asymmetrically within the computational domain (upper interface at $z=0$), so that changing $F$ also changes the absolute potential reference across the domain.
The conventional local model therefore exhibits an unphysical asymmetry under $F\to -F$, reflecting its energy-reference dependence.
We note that the unphysical reference energy dependence here is intertwined with a spurious coordinate dependence.
In contrast, both the exact non-local model and the projected-local model obey the required symmetry and agree quantitatively over the field range considered.

\subsubsection{Wiggle Well}

\begin{figure}
\includegraphics[width=1\columnwidth]{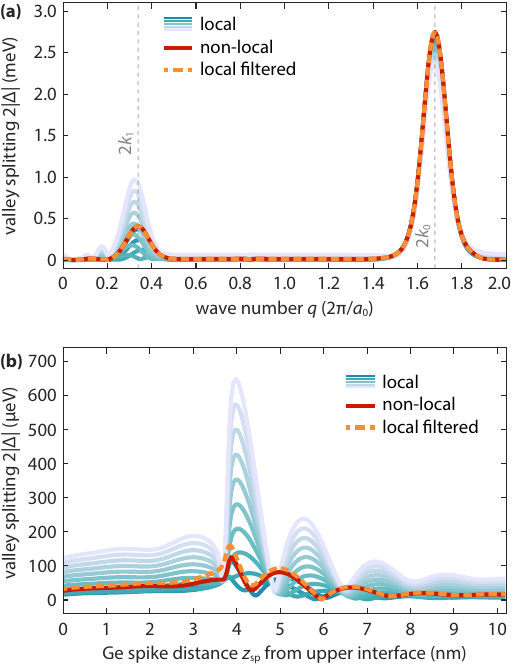}
\caption{Valley splitting in unconventional heterostructures.
\textbf{(a)}~Valley splitting in a wiggle well with Ge amplitude $X_{\mathrm{ww}}=0.05$ and varying wave number $q$.
The dominant resonances at $q=2k_{0}$ and $q=2k_{1}$ are captured by both the local and the non-local model, but the quantitative results near $q=2k_{1}$ show a strong unphysical reference energy dependence when using the local model.
\textbf{(b)}~Numerical results for a Ge spike with amplitude $x_{s}=0.6$ and width $\sigma_{s}=2\,\mathrm{ML}$.
The predictions of local and non-local theory differ significantly due to the sharp feature in the mesoscopic potential, which leads to unphysical contributions from short-wavelength components in the local EFT.
The projected-local envelope approach is in good qualitative agreement with the exact result from the non-local model, but it tends to overestimate the actual valley splitting. 
\label{fig: unconventional}
}
\end{figure}

Oscillating Ge concentration profiles within the QW, known as \emph{wiggle wells}, can lead to strong enhancements of the intervalley coupling matrix element when the modulation wave number is resonant with
the valley wave-vector separation \cite{Feng2022, McJunkin2022, Woods2024, Thayil2025, Gradwohl2025, Thayil2025c, Cvitkovich2026}.
Within the continuum approach followed here, the wiggle well is described by a Ge concentration profile of the form $X\left(z\right) = X_{\mathrm{QW}}\left(z\right)+X_{\mathrm{mod}}\left(z\right)$ with
\begin{equation}
X_{\mathrm{mod}}\left(z\right) =\frac{1}{2}X_{\mathrm{ww}}\left(1+\cos\left(qz\right)\right)\,\Xi\left(z\right),\label{eq: wiggle well}
\end{equation}
where $X_{\mathrm{ww}}$ is the Ge concentration amplitude, $q$ is the wiggle well wave number and $\Xi\left(z\right)$ is the QW shape function, see Eq.~(\ref{eq: QW indicator}).
Numerical results for
a wiggle well profile with varying wave number $q$ are shown in Fig.~\ref{fig: unconventional}\,(a).
Both the local and the non-local models capture the strong enhancement at $q=2k_{0}$ (``short-period wiggle well'', period length
$\lambda=\pi/k_{0}\approx2.4\,\mathrm{ML}$), which corresponds to resonant coupling of valley states within the same Brillouin zone, cf. Fig.~\ref{fig:grid-brillouin}(a).
A second resonance appears at $q=2k_{1}$, where
\begin{equation}
    k_{1}=2\pi(1-\varepsilon_{z,z})/a_{0}-k_{0} \label{eq: k1}
\end{equation}
is the distance from the valley minimum to the Brillouin zone edge.
This ``long-period wiggle well'' has a period length of $\lambda=\pi/k_{1}\approx11.8\,\mathrm{ML}$ and triggers a resonance between valley states in neighboring
Brillouin zones.
The underlying resonance mechanism, however, is blocked
in highly symmetric silicon lattices and must be unlocked by either alloy disorder \cite{Feng2022,Cvitkovich2026} or shear strain \cite{Woods2024,Thayil2025}.
Near this $2 k_1$-resonance, the conventional local model exhibits a pronounced energy-reference dependence, whereas the projected-local model closely tracks the exact non-local result.

\subsubsection{Germanium Spike}
Finally, we consider a Ge spike inside the QW \cite{McJunkin2021, Losert2023, Salamone2025}, modeled by $X\left(z\right) = X_{\mathrm{QW}}\left(z\right)+X_{\mathrm{mod}}\left(z\right)$
with a Gaussian spike
\begin{equation}
X_{\mathrm{mod}}\left(z\right) =X_{\mathrm{sp}}\,\exp\left(-\frac{1}{2}\left(\frac{z-z_{\mathrm{sp}}}{\sigma_{\mathrm{sp}}}\right)^{2}\right),\label{eq: Ge spike}
\end{equation}
where $X_{\mathrm{sp}}$ is the peak Ge concentration, $z_{\mathrm{sp}}$ is the spike position
and $\sigma_{\mathrm{sp}}$ is the spike width.
Figure~\ref{fig: unconventional}\,(b) shows the resulting valley splitting as a function of the spike position $z_{\mathrm{sp}}$.
Because the spike introduces a sharp feature in $U_{\perp}\left(z\right)$, especially at high Ge peak concentrations, the local model can acquire substantial short-wavelength components and exhibits strong energy-reference dependence.
The projected-local approximation substantially reduces this artifact but tends to overestimate the exact non-local result, indicating that simple projection does not capture all aspects of the non-local model.

\section{Summary and Conclusion}
Accurate modeling of valley splitting in Si/SiGe nanostructures is essential for assessing and designing heterostructures for spin-qubit devices.
Conventional local envelope function theory (EFT) is computationally efficient, but for multi-valley semiconductors it implicitly requires that the envelope wave functions remain confined to valley-specific sectors of the Brillouin zone.
In heterostructures with atomically sharp interfaces or engineered Ge profiles, this band-limitation is generically violated in the local model, leading to spectral leakage between valley sectors.
A particularly severe consequence is that the intervalley coupling matrix element obtained from the local theory acquires an unphysical dependence on the choice of energy reference, \emph{i.e.}, it is not invariant under global potential offsets.
This energy-reference dependence renders the valley splitting predicted by the local model non-unique, complicating both quantitative prediction and the interpretation of experimental data in sharply varying potential landscapes.

In this work, we developed an exact multi-valley EFT of Burt--Foreman type that treats the valley-sector projection exactly and yields a non-local envelope equation for the band-limited envelopes.
Within this framework, the intervalley coupling and the valley splitting are uniquely defined and strictly invariant under global shifts of the confinement potential.
We have quantified the magnitude of the unphysical ambiguity of the local envelope model by numerical simulations of a number of Si/SiGe heterostructure designs and benchmarked a simple projected-local (spectrally filtered) approximation.
The projected-local approach restores the energy-reference invariance and approximates the exact non-local results well in most cases.

\section*{Data Availability}

MATLAB code to reproduce the simulation results and figures of this
work is available on GitHub \cite{Ermoneit2026b}.

\begin{acknowledgments}
This work was funded by the Deutsche Forschungsgemeinschaft (DFG,
German Research Foundation) under Germany's Excellence Strategy --
The Berlin Mathematics Research Center MATH+ (EXC-2046/1, project
ID: 390685689).
\end{acknowledgments}

\appendix

\section{Justification of the Expansion Ansatz \label{sec: ansatz justification}}

We consider an expansion of the microscopic wave function in Bloch
waves (linear combination of bulk eigenstates)
\begin{equation}
\Psi_{\alpha}\left(\mathbf{r}\right) =\sum_{n}\sum_{\mathbf{k}\in\mathrm{FBZ}}a_{n,\mathbf{k},\alpha}\phi_{n,\mathbf{k}}\left(\mathbf{r}\right)\label{eq: expansion in Bloch waves}
\end{equation}
 where the Bloch wave function reads
\begin{equation*}
\phi_{n,\mathbf{k}}\left(\mathbf{r}\right) =\frac{1}{\sqrt{V_{c}}}\mathrm{e}^{\mathrm{i}\mathbf{k}\cdot\mathbf{r}}u_{n,\mathbf{k}}\left(\mathbf{r}\right).
\end{equation*}
The summation over plane wave components is restricted to the first
Brillouin zone (FBZ).
The Bloch factors $u_{n,\mathbf{k}}\left(\mathbf{r}\right)$ are
complete and orthogonal on the space of lattice-periodic functions, see Eqs.~(\ref{eq:u-complete})--(\ref{eq:u-orthonormal}).
The Bloch waves satisfy
\begin{align*}
\int_{V_{c}}\mathrm{d}^{3}r\,\phi_{n,\mathbf{k}}^{*}\left(\mathbf{r}\right)\phi_{n',\mathbf{k}'}\left(\mathbf{r}\right) & =\delta_{n,n'}\delta_{\mathbf{k},\mathbf{k}'},\\
\sum_{n,\mathbf{k}}\phi_{n,\mathbf{k}}\left(\mathbf{r}\right)\phi_{n,\mathbf{k}}^{*}\left(\mathbf{r}'\right) & =\delta\left(\mathbf{r}-\mathbf{r}'\right).
\end{align*}
Along with the requirements on the expansion coefficients $\sum_{n}\sum_{\mathbf{k}\in\mathrm{FBZ}}a_{n,\mathbf{k},\alpha}^{*}a_{n,\mathbf{k},\alpha'}=\delta_{\alpha,\alpha'}$
and $\sum_{\alpha}a_{n,\mathbf{k},\alpha}a_{n',\mathbf{k}',\alpha}^{*}=\delta_{n,n'}\delta_{\mathbf{k},\mathbf{k}'}$,
this guarantees complete orthonormality of the microscopic wave function
(\ref{eq:psi-orthonormal})--(\ref{eq:psi-complete}).

For multi-valley systems it is convenient to rewrite (\ref{eq: expansion in Bloch waves})
using the decomposition of the Brillouin zone into non-overlapping valley-specific
sectors
\begin{equation*}
\mathrm{FBZ} =\bigcup_{\mathbf{k}_{0}}S\left(\mathbf{k}_{0}\right),
\end{equation*}
see Fig.~\ref{fig:grid-brillouin}\,(b), as
\begin{align}
\Psi_{\alpha}\left(\mathbf{r}\right) & =\sum_{n,\mathbf{k}_{0}}\sum_{\mathbf{K}\in S\left(\mathbf{k}_{0}\right)}a_{n,\mathbf{K},\alpha}\phi_{n,\mathbf{K}}\left(\mathbf{r}\right)\label{eq: expansion of Bloch waves - FBZ decompostion}\\
 & =\sum_{n,\mathbf{k}_{0}}\sum_{\mathbf{K}\in S\left(\mathbf{k}_{0}\right)}a_{n,\mathbf{K},\alpha}\frac{1}{\sqrt{V_{c}}}\mathrm{e}^{\mathrm{i}\mathbf{K}\cdot\mathbf{r}}u_{n,\mathbf{K}}\left(\mathbf{r}\right),\nonumber
\end{align}
where $\mathbf{K}$ is the crystal momentum in the respective valley-sector (not shifted).
Using the completeness of Bloch factors (\ref{eq:u-complete}), we
write Eq.~(\ref{eq: expansion of Bloch waves - FBZ decompostion})
as
\begin{align*}
\Psi_{\alpha}\left(\mathbf{r}\right) & =\sum_{n,\mathbf{k}_{0}}\sum_{\mathbf{K}\in S\left(\mathbf{k}_{0}\right)}a_{n,\mathbf{K},\alpha}\frac{1}{\sqrt{V_{c}}}\mathrm{e}^{\mathrm{i}\mathbf{K}\cdot\mathbf{r}}\times\\
 & \phantom{=}\times\int_{\Omega_{p}}\mathrm{d}^{3}r'\,\underbrace{\frac{1}{\Omega_{p}}\sum_{m}u_{m,\mathbf{k}_{0}}\left(\mathbf{r}\right)u_{m,\mathbf{k}_{0}}^{*}\left(\mathbf{r}'\right)}_{=\delta\left(\mathbf{r}-\mathbf{r}'\right)}u_{n,\mathbf{K}}\left(\mathbf{r}'\right)\\
 & =\sum_{m,\mathbf{k}_{0}}u_{m,\mathbf{k}_{0}}\left(\mathbf{r}\right)\sum_{\mathbf{K}\in S\left(\mathbf{k}_{0}\right)}\mathrm{e}^{\mathrm{i}\mathbf{K}\cdot\mathbf{r}}F_{m,\mathbf{k}_{0},\alpha}\left(\mathbf{K}\right)
\end{align*}
with the Fourier domain envelope wave function 
\begin{equation*}
F_{m,\mathbf{k}_{0},\alpha}\left(\mathbf{K}\right) =\frac{1}{\sqrt{V_{c}}}\sum_{n}a_{n,\mathbf{K},\alpha}
\int_{\Omega_{p}}\frac{\mathrm{d}^{3}r'}{\Omega_{p}}\,u_{m,\mathbf{k}_{0}}^{*}\left(\mathbf{r}'\right)u_{n,\mathbf{K}}\left(\mathbf{r}'\right).
\end{equation*}
Using inverse Fourier transform
\begin{equation*}
F_{m,\mathbf{k}_{0},\alpha}\left(\mathbf{r}\right) =\sum_{\mathbf{K}\in S\left(\mathbf{k}_{0}\right)}\mathrm{e}^{\mathrm{i}\mathbf{K}\cdot\mathbf{r}}F_{m,\mathbf{k}_{0},\alpha}\left(\mathbf{K}\right),
\end{equation*}
we arrive at the ansatz (\ref{eq:decomposition}) used in the main text.
The properties (\ref{eq: F-orthonormal})--(\ref{eq: F-complete}) of the envelope wave function $F_{n,\mathbf{k}_{0},\alpha}$ follow directly from the considerations above. 
The restriction $K\in S\left(\mathbf{k}_0\right)$ prevents overcompleteness and is the origin of the sector-projected delta function $\Delta_{\mathbf{k}_0}$ used in the exact theory.

\section{Proof of Gauge Invariance \label{sec: proof gauge invariance}}

We consider $U\left(\mathbf{r}\right)\to U\left(\mathbf{r}\right)+U_{0}$.
Using (\ref{eq: mesoscopic potential matrix element U}) we obtain
\begin{equation*}
U_{\mathbf{k}_{0},\mathbf{k}_{0}'}^{n,n'}\left(\mathbf{r},\mathbf{r}'\right)
\xrightarrow{U\to U+U_{0}}  U_{\mathbf{k}_{0},\mathbf{k}_{0}'}^{n,n'}\left(\mathbf{r},\mathbf{r}'\right)+T_{\mathbf{k}_{0},\mathbf{k}_{0}'}^{n,n'}
\end{equation*}
where the new term involving the constant $U_{0}$ is
\begin{align*}
T_{\mathbf{k}_{0},\mathbf{k}_{0}'}^{n,n'}\left(\mathbf{r},\mathbf{r}'\right) & =U_{0}\int_{V_{c}}\mathrm{d}^{3}r''\,\Delta_{\mathbf{k}_{0}}\left(\mathbf{r}-\mathbf{r}''\right)\\
 & \times u_{n,\mathbf{k}_{0}}^{*}\left(\mathbf{r}''\right)u_{n',\mathbf{k}_{0}'}\left(\mathbf{r}''\right)\Delta_{\mathbf{k}_{0}'}\left(\mathbf{r}''-\mathbf{r}'\right).
\end{align*}
The new term is evaluated in Fourier domain using Eqs.~(\ref{eq: Bloch Fourier series}) and (\ref{eq: Delta low pass}) as
\begin{align*}
&T_{\mathbf{k}_{0},\mathbf{k}_{0}'}^{n,n'} \left(\mathbf{r},\mathbf{r}'\right)=U_{0}\,\frac{1}{V_{c}}\sum_{\mathbf{K}\in S\left(\mathbf{k}_{0}\right)}\sum_{\mathbf{K}'\in S\left(\mathbf{k}_{0}'\right)}\mathrm{e}^{\mathrm{i}\mathbf{K}\cdot\mathbf{r}}\mathrm{e}^{-\mathrm{i}\mathbf{K}'\cdot\mathbf{r}'}\times\\
&\times \sum_{\mathbf{G},\mathbf{G}'}c_{n,\mathbf{k}_{0}}^{*}\left(\mathbf{G}\right)c_{n',\mathbf{k}_{0}'}\left(\mathbf{G}'\right)\underbrace{\frac{1}{V_{c}}\int_{V_{c}}\mathrm{d}^{3}r''\,\mathrm{e}^{-\mathrm{i}\left(\mathbf{G}-\mathbf{G}'+\mathbf{K}-\mathbf{K}'\right)\cdot\mathbf{r}''}}_{=\delta_{\mathbf{G}-\mathbf{G}'+\mathbf{K}-\mathbf{K}',\mathbf{0}}}.
\end{align*}
Carrying out the summation over $\mathbf{K}'$ yields
\begin{align*}
T_{\mathbf{k}_{0},\mathbf{k}_{0}'}^{n,n'}\left(\mathbf{r},\mathbf{r}'\right) & =U_{0}\frac{1}{V_{c}}\sum_{\mathbf{K}}\sum_{\mathbf{G},\mathbf{G}'}\chi_{S\left(\mathbf{k}_{0}\right)}\left(\mathbf{K}\right)\chi_{S\left(\mathbf{k}_{0}'\right)}\left(\mathbf{G}-\mathbf{G}'+\mathbf{K}\right)\\
 & \times c_{n,\mathbf{k}_{0}}^{*}\left(\mathbf{G}\right)c_{n',\mathbf{k}_{0}'}\left(\mathbf{G}'\right)\mathrm{e}^{\mathrm{i}\mathbf{K}\cdot\left(\mathbf{r}-\mathbf{r}'\right)}\mathrm{e}^{-\mathrm{i}\left(\mathbf{G}-\mathbf{G}'\right)\cdot\mathbf{r}'}
\end{align*}
where we used the indicator function
\begin{equation*}
\chi_{S\left(\mathbf{k}_{0}\right)}\left(\mathbf{K}\right) =\begin{cases}
1 & \mathbf{K}\in S\left(\mathbf{k}_{0}\right),\\
0 & \mathrm{else}
\end{cases}
\end{equation*}
that is non-zero only for plane wave components from the corresponding
Brillouin zone sector. Since the combination of reciprocal lattice
vectors $\left(\mathbf{G}-\mathbf{G}'\right)$ is within the first Brillouin
zone only for $\mathbf{G}=\mathbf{G}'$, we arrive at
\begin{align*}
T_{\mathbf{k}_{0},\mathbf{k}_{0}'}^{n,n'}\left(\mathbf{r},\mathbf{r}'\right) & =U_{0}\frac{1}{V_{c}}\sum_{\mathbf{K}}\chi_{S\left(\mathbf{k}_{0}\right)}\left(\mathbf{K}\right)\chi_{S\left(\mathbf{k}_{0}'\right)}\left(\mathbf{K}\right)\mathrm{e}^{\mathrm{i}\mathbf{K}\cdot\left(\mathbf{r}-\mathbf{r}'\right)}\\
 & \times\sum_{\mathbf{G}}c_{n,\mathbf{k}_{0}}^{*}\left(\mathbf{G}\right)c_{n',\mathbf{k}_{0}'}\left(\mathbf{G}\right)
\end{align*}
Since the Brillouin zone sectors are disjoint $S\left(\mathbf{k}_{0}\right)\cap S\left(\mathbf{k}_{0}'\right)=\emptyset$
for $\mathbf{k}_{0}\neq\mathbf{k}_{0}'$, it holds
\begin{equation*}
\chi_{S\left(\mathbf{k}_{0}\right)}\left(\mathbf{K}\right)\chi_{S\left(\mathbf{k}_{0}'\right)}\left(\mathbf{K}\right) \equiv\delta_{\mathbf{k}_{0},\mathbf{k}_{0}'}\chi_{S\left(\mathbf{k}_{0}\right)}\left(\mathbf{K}\right).
\end{equation*}
Using the orthogonality relation
\begin{equation*}
\sum_{\mathbf{G}}c_{n,\mathbf{k}_{0}}^{*}\left(\mathbf{G}\right)c_{n',\mathbf{k}_{0}}\left(\mathbf{G}\right)  =\delta_{n,n'},
\end{equation*}
we arrive at
\begin{equation*}
T_{\mathbf{k}_{0},\mathbf{k}_{0}'}^{n,n'}\left(\mathbf{r},\mathbf{r}'\right) =U_{0}\,\delta_{n,n'}\delta_{\mathbf{k}_{0},\mathbf{k}_{0}'}\Delta_{\mathbf{k}_{0}}\left(\mathbf{r}-\mathbf{r}'\right)
\end{equation*}
which implies the identity (\ref{eq: gauge transform u}) stated in the main text.

\section{Effective One-Dimensional Schrödinger Equation \label{sec: effective one-dim Schroedinger}}

\begin{table}[t]
\begin{tabular*}{1\columnwidth}{@{\extracolsep{\fill}}ccc}
\toprule 
$n$ & $C_{n}^{\left(2\right)}$ & $B_{n}$\tabularnewline
\midrule
$-4$ & $1.84\times10^{-3}$ & $B_{4}$\tabularnewline
$-3$ & $0.599\cdot\varepsilon_{x,y}$ & $B_{3}$\tabularnewline
$-2$ & $-2.44\times10^{-2}$ & $B_{2}$\tabularnewline
$-1$ & $-31.8\cdot\varepsilon_{x,y}$ & $B_{1}$\tabularnewline
$\phantom{+}0$ & $-0.221$ & $1.0$\tabularnewline
$\phantom{+}1$ & $-0.402\cdot\varepsilon_{x,y}$ & $\phantom{+}2.92\cdot\varepsilon_{x,y}$\tabularnewline
$\phantom{+}2$ & $1.58\times10^{-3}$ & $-5.79\times10^{-4}$\tabularnewline
$\phantom{+}3$ & $1.04\times10^{-2}\cdot\varepsilon_{x,y}$ & $-8.55\times10^{-2}\cdot\varepsilon_{x,y}$\tabularnewline
$\phantom{+}4$ & $3.35\times10^{-5}$ & $-2.47\times10^{-4}$\tabularnewline
\bottomrule
\end{tabular*}

\caption{Leading coefficients $C_{n}^{\left(2\right)}$ contributing to the valley splitting, see Eq.~(\ref{eq: C2 coefficient}), and $B_{n}$ to the effective one-dimensional
unperturbed single-valley problem, see Eq.~(\ref{eq: B coefficient}).
The coefficients were computed using empirical pseudopotential theory \cite{Thayil2025} for a strained silicon crystal with strain tensor (\ref{eq: strain tensor}).
The linear dependence of odd-integer coefficients on shear strain was obtained from a least-squares fit to numerical data.
\label{tab:Bloch coefficients}
}
\end{table}

We consider the zeroth-order single-valley problem (\ref{eq: zeroth order H_0^+})
\begin{align}
E_{\alpha}f_{\mathbf{k}_{0}^{+},\alpha}^{\left(0\right)}\left(\mathbf{r}\right) & =\left(-\frac{\hbar^{2}}{2}\nabla\cdot\left(m_{c}^{-1}\nabla\right)+E_{c}\right)f_{\mathbf{k}_{0}^{+},\alpha}^{\left(0\right)}\left(\mathbf{r}\right)\label{eq: single valley eigenvalue problem f+}\\
 & \phantom{=}+\int_{V_{c}}\mathrm{d}^{3}r'\,u_{\mathbf{k}_{0}^{+},\mathbf{k}_{0}^{+}}\left(\mathbf{r},\mathbf{r}'\right)f_{\mathbf{k}_{0}^{+},\alpha}^{\left(0\right)}\left(\mathbf{r}'\right)\nonumber 
\end{align}
arising in degenerate perturbation theory.
The ground state slowly varying envelope $f_{\mathbf{k}_{0}^{+}}^{\left(0\right)}=\left(f_{\mathbf{k}_{0}^{-}}^{\left(0\right)}\right)^{*}$ enters the intervalley coupling matrix element in first-order perturbation theory.
In Eq.~(\ref{eq: single valley eigenvalue problem f+}), the intravalley kernel \(u_{\mathbf{k}_{0}^{+},\mathbf{k}_{0}^{+}}\) contains rapidly oscillating Bloch factors, which weakly break the symmetry of the mesoscopic potential $U$.
As a result, the envelope is not strictly separable.
To obtain a tractable model for gate-defined QDs with lateral length scales $l_{x,y}\gg a_{0}$, we (i)~approximate separability into in-plane and vertical components, (ii)~integrate out the transverse coordinates using the harmonic-oscillator ground state, and (iii)~exploit the valley-sector projection to formulate the resulting one-dimensional problem in Fourier space on the admitted Brillouin zone segment.

Since the in-plane confinement varies slowly, we impose the separation ansatz
\begin{equation}
f_{\mathbf{k}_{0}^{+},\alpha}^{(0)}(\mathbf r)\approx \phi_{0}(x,y)\,f_{\mathbf{k}_{0}^{+},\alpha}^{(0)}(z)
\label{eq: separation ansatz for 1D problem}
\end{equation}
with the ground state wave function of the two-dimensional harmonic oscillator 
\begin{equation*}
\phi_{0}\left(x,y\right) =\frac{1}{\pi^{1/4}l_{x}^{1/2}}\mathrm{e}^{-\frac{1}{2}\left(x/l_{x}\right)^{2}}\times\frac{1}{\pi^{1/4}l_{y}^{1/2}}\mathrm{e}^{-\frac{1}{2}\left(y/l_{y}\right)^{2}},
\end{equation*}
which obey
\begin{align}
 & \frac{1}{2}\left(\hbar\omega_{x}+\hbar\omega_{y}\right)\phi_{0}\left(x,y\right)=\label{eq: 2d harmonic oscillator}\\
 & \qquad=\left[-\frac{\hbar^{2}}{2m_{t}}\left(\frac{\partial^{2}}{\partial x^{2}}+\frac{\partial^{2}}{\partial y^{2}}\right)+U_{\text{QD}}\left(x,y\right)\right]\phi_{0}\left(x,y\right).\nonumber 
\end{align}
We assume that the QD is much larger than the atomic unit cell $l_{x,y}\gg a_{0}$, where $l_{x,y}=\left(\hbar/m_{t}\omega_{x,y}\right)^{1/2}$.
In this limit, we approximate the truncated delta function as
\begin{equation*}
\Delta_{\mathbf{k}_{0}}\left(\mathbf{r}-\mathbf{r}'\right) \approx\delta\left(x-x'\right)\delta\left(y-y'\right)\Delta_{k_{0}}\left(z-z'\right),
\end{equation*}
since the restriction to the valley-sectors is only relevant in the growth direction of the heterostructure via
\begin{equation*}
\Delta_{k_{0}}\left(z-z'\right) =\int_{S_{+}}\frac{\mathrm{d}K}{2\pi}\,\mathrm{e}^{\mathrm{i}K\left(z-z'\right)},
\end{equation*}
where ${S_{+}=\left(0,\frac{2\pi}{a_{0}}\left[1-\varepsilon_{z,z}\right]\right)}$ is the one-dimensional Brillouin zone segment.
Here we consider a strain-induced modification of the Brillouin zone edge due to biaxial strain arising in pseudomorphic lattice-matched growth of the SiGe heterostructure.
In addition, we assume a weak shear strain $\varepsilon_{x,y}$
within the QW layer along the {[}110{]} direction. The full strain
tensor reads \cite{Thayil2025}
\begin{equation}
\varepsilon =\left(\begin{array}{ccc}
\varepsilon_{\parallel} & \varepsilon_{x,y} & 0\\
\varepsilon_{x,y} & \varepsilon_{\parallel} & 0\\
0 & 0 & \varepsilon_{\perp}
\end{array}\right).\label{eq: strain tensor}
\end{equation}
Integrating (\ref{eq: single valley eigenvalue problem f+}) with
$\phi_{0}^{*}\left(x,y\right)$ over the transverse plane and using
(\ref{eq: separation ansatz for 1D problem}) and (\ref{eq: 2d harmonic oscillator})
yields 
\begin{align*}
E_{\alpha}F_{\mathbf{k}_{0}^{+},\alpha}^{\left(0\right)}\left(z\right) & =\left(-\frac{\hbar^{2}}{2m_{l}}\frac{\partial^{2}}{\partial z^{2}}+E_{c}+\frac{1}{4}\left(\hbar\omega_{x}+\hbar\omega_{y}\right)\right)F_{\mathbf{k}_{0}^{+},\alpha}^{\left(0\right)}\left(z\right)\\
 & \phantom{=}+\int\mathrm{d}z'\,\Delta_{k_{0}^{+}}\left(z-z'\right)U_{\mathrm{eff}}\left(z'\right)F_{\mathbf{k}_{0}^{+},\alpha}^{\left(0\right)}\left(z'\right)
\end{align*}
with $F_{\mathbf{k}_{0}^{+},\alpha}^{\left(0\right)}\left(z\right)=\mathrm{e}^{\mathrm{i}k_{0}z}f_{\mathbf{k}_{0}^{+},\alpha}^{\left(0\right)}\left(z\right)$
and the effective potential
\begin{align*}
&U_{\mathrm{eff}}\left(z\right)=\sum_{\mathbf{G},\mathbf{G}'}c_{\mathbf{k}_{0}^{+}}^{*}\left(\mathbf{G}\right)c_{\mathbf{k}_{0}^{+}}\left(\mathbf{G}'\right)\mathrm{e}^{-\big(\frac{[G_{x}-G_{x}']l_{x}}{2}\big)^{2}}\mathrm{e}^{-\big(\frac{[G_{y}-G_{y}']l_{y}}{2}\big)^{2}}\\
&\hphantom{=}\times\mathrm{e}^{-\mathrm{i}\left(G_{z}-G_{z}'\right)z}\,\bigg(U_{\perp}\left(z\right)+\frac{\hbar\omega_{x}}{2}\bigg(\frac{1}{2}-\bigg[\frac{\big(G_{x}-G_{x}'\big)l_{x}}{2}\bigg]^{2}\bigg)\\
&\hphantom{=\times\mathrm{e}^{-\mathrm{i}\left(G_{z}-G_{z}'\right)z}\,\bigg(}+\frac{\hbar\omega_{y}}{2}\bigg(\frac{1}{2}-\bigg[\frac{\big(G_{y}-G_{y}'\big)l_{y}}{2}\bigg]^{2}\bigg)\bigg).
\end{align*}
At small strain, the reciprocal lattice vectors read
\begin{equation*}
\mathbf{G}-\mathbf{G}' \approx \left(I-\varepsilon\right)\sum_{i=1,2,3}\Delta n_{i}\mathbf{b}_{i},
\end{equation*}
where $\Delta n_{i}\in\mathbb{Z}$, $i=\left\{ 1,2,3\right\} $ and
$\mathbf{b}_{i}$ are the primitive reciprocal lattice vectors of
the relaxed fcc lattice
\begin{align*}
\mathbf{b}_{1} & =\frac{2\pi}{a_{0}}\left(\begin{array}{c}
-1\\
+1\\
+1
\end{array}\right), & \mathbf{b}_{2} & =\frac{2\pi}{a_{0}}\left(\begin{array}{c}
+1\\
-1\\
+1
\end{array}\right), & \mathbf{b}_{3} & =\frac{2\pi}{a_{0}}\left(\begin{array}{c}
+1\\
+1\\
-1
\end{array}\right).
\end{align*}
For small strain and large QDs, the Gaussians strongly suppress most reciprocal lattice vector combinations except for \cite{Thayil2025}
\begin{subequations}
\label{eq: Kronecker selection rule}
\begin{align}
\mathrm{e}^{-\big(\frac{[G_{x}-G_{x}']l_{x}}{2}\big)^{2}}
&\approx\delta_{-\Delta n_{1}+\Delta n_{2}+\Delta n_{3},0}
\,\mathrm{e}^{-\big(\frac{[G_{x}-G_{x}']l_{x}}{2}\big)^{2}},\\
\mathrm{e}^{-\big(\frac{[G_{y}-G_{y}']l_{y}}{2}\big)^{2}}
& \approx\delta_{+\Delta n_{1}-\Delta n_{2}+\Delta n_{3},0}
\,\mathrm{e}^{-\big(\frac{[G_{y}-G_{y}']l_{y}}{2}\big)^{2}},
\end{align}
\end{subequations}
Using the selection rules~(\ref{eq: Kronecker selection rule}), we arrive at
\begin{align*}
U_{\mathrm{eff}}\left(z\right) & =\sum_{n\in\mathbb{Z}}B_{n}\mathrm{e}^{-\big(\frac{1}{2}nG_{0,x}l_{x}\big)^{2}}\mathrm{e}^{-\big(\frac{1}{2}nG_{0,y}l_{y}\big)^{2}}\mathrm{e}^{-\mathrm{i}nG_{0,z}z}\times\\
 & \phantom{=}\times\bigg(U_{\perp}\left(z\right)+\frac{\hbar\omega_{x}}{2}\bigg(\frac{1}{2}-\bigg[\frac{nG_{0,x}l_{x}}{2}\bigg]^{2}\bigg)\\
 & \phantom{=\times\bigg(}+\frac{\hbar\omega_{y}}{2}\bigg(\frac{1}{2}-\bigg[\frac{nG_{0,y}l_{y}}{2}\bigg]^{2}\bigg)\bigg).
\end{align*}
where we introduced the vector
\begin{equation}
\mathbf{G}_{0}=\left(I-\varepsilon\right)\left(\mathbf{b}_{1}+\mathbf{b}_{2}\right)=\frac{4\pi}{a_{0}}\left(\begin{array}{c}
-\varepsilon_{z,x}\\
-\varepsilon_{y,z}\\
1-\varepsilon_{z,z}
\end{array}\right)\label{eq: G0 vector}
\end{equation}
and the band-structure coefficients
\begin{equation}
B_{n}=\sum_{\mathbf{G},\mathbf{G}'}c_{\mathbf{k}_{0}^{+}}^{*}\left(\mathbf{G}\right)c_{\mathbf{k}_{0}^{+}}\left(\mathbf{G}'\right)\delta_{\mathbf{G}-\mathbf{G}',\mathbf{G}_{0}n}.\label{eq: B coefficient}
\end{equation}
The leading term is $B_{0}=1$ (normalization of Bloch factors) and
it holds $B_{n}^{*}=B_{-n}$. Numerical data is given in Tab.~\ref{tab:Bloch coefficients}.
Transformation to the Fourier domain yields 
\begin{align*}
E_{\alpha}F_{\mathbf{k}_{0}^{+},\alpha}^{\left(0\right)}\left(K\right)
&=
\left(\frac{\hbar^{2}K^{2}}{2m_{l}}+E_{c}+\frac{1}{4}\left(\hbar\omega_{x}+\hbar\omega_{y}\right)\right)
F_{\mathbf{k}_{0}^{+},\alpha}^{\left(0\right)}\left(K\right)\\
&\phantom{=}+\sum_{K'\in S_{+}}U_{\mathrm{eff}}\left(K-K'\right)F_{\mathbf{k}_{0}^{+},\alpha}^{\left(0\right)}\left(K'\right),
\end{align*}
with the Fourier domain effective potential 
\begin{align*}
U_{\mathrm{eff}} & \left(K-K'\right)=\frac{1}{4}\left(\hbar\omega_{x}+\hbar\omega_{y}\right)\delta_{K,K'}+\\
+ & \sum_{n\in\mathbb{Z}}B_{n}\mathrm{e}^{-\left(\frac{1}{2}nG_{0,x}l_{x}\right)^{2}}\mathrm{e}^{-\left(\frac{1}{2}nG_{0,y}l_{y}\right)^{2}}U_{\perp}\left(K-K'+nG_{0,z}\right).
\end{align*}
In the last step we used the identity $\delta_{K-K'+nG_{0,z},0}=\delta_{K,K'}\delta_{n,0}$
that holds for $K,K'\in S_{+}=\left(0,G_{0,z}/2\right)$.
The $\delta_{K,K'}$ term contributes an additional $\frac{1}{4}\left(\hbar\omega_{x}+\hbar\omega_{y}\right)$ to the diagonal, so that the total harmonic-oscillator zero-point energy becomes  $\frac{1}{2}\left(\hbar\omega_{x}+\hbar\omega_{y}\right)$ in Eq.~(\ref{eq: effective Schroedinger equation Fourier space (appendix)}).
We note that short-wavelength components of the potential $U_{\perp}$ enter via backfolding to the Brillouin zone sector with coefficients $B_{n\neq0}$.
Shear strain components entering $G_{0,x}$ and $G_{0,y}$ lead to additional Gaussian damping.
For the sake of simplicity, we assume $\varepsilon_{z,x}=\varepsilon_{y,z}=0$ throughout this work, such that we arrive at
\begin{align}
E_{\alpha}F_{\mathbf{k}_{0}^{+},\alpha}^{\left(0\right)}\left(K\right) & =\left(\frac{\hbar^{2}K^{2}}{2m_{l}}+E_{c}+\frac{1}{2}\left(\hbar\omega_{x}+\hbar\omega_{y}\right)\right)F_{\mathbf{k}_{0}^{+},\alpha}^{\left(0\right)}\left(K\right)\nonumber \\
 & +\sum_{K'\in S_{+}}\sum_{n\in\mathbb{Z}}B_{n}U_{\perp}\left(K-K'+nG_{0,z}\right)F_{\mathbf{k}_{0}^{+},\alpha}^{\left(0\right)}\left(K'\right).
 \label{eq: effective Schroedinger equation Fourier space (appendix)}
\end{align}
Equation (\ref{eq: effective Schroedinger equation Fourier space (appendix)})
is solved numerically for $K\in S_{+}$, see Fig.~\ref{fig: wave function}\,(b).
The envelope wave function of the opposite valley state is obtained from the identity
$F_{\mathbf{k}_{0}^{-},\alpha}^{\left(0\right)}\left(K\right) =\left(F_{\mathbf{k}_{0}^{+},\alpha}^{\left(0\right)}\left(-K\right)\right)^{*}$
for $K\in S_{-}$ with $S_{-}=\left(-G_{0,z}/2,0\right)$.

\section{Numerical Method \label{sec: numerical method}}

For the numerical solution of Eq.~(\ref{eq: effective Schroedinger equation Fourier space})
we use an equidistant grid with $N$ points constructed in
Fourier space with total number of points $N=N_{\mathrm{BZ}}\cdot N_{\mathrm{FBZ}}.$
The number of resolved Brillouin zones $N_{\mathrm{BZ}}$ determines
the spatial resolution $\Delta z=2\pi/\left(G_{0,z}N_{\mathrm{BZ}}\right)$
and the number of points in the first Brillouin zone $N_{\mathrm{FBZ}}$
determines the domain size $L=2\pi N_{\mathrm{FBZ}}/G_{0,z}$. The
Brillouin zone edges are part of the grid by construction,
which is required to construct the indicator functions $S_{\pm}$.
The valley minima at $\pm k_{0}$, however, are in general not part of the grid and require sufficiently high resolution
(large $N_{\mathrm{FBZ}}$).

The kinetic operator is discretized using the spectral method. The mesoscopic potential $U_{\perp}\left(z\right)$ is constructed on the corresponding position space grid first and then transformed to Fourier domain.
Short-wavelength components are backfolded to the Brillouin zone sectors using the coefficients $B_{n}$.
We assume periodic boundary conditions.

The resulting matrix eigenvalue problem is solved for the admissible wave numbers ${K\in S_+}$ (or $S_{-}$) from the selected Brillouin zone segment to enforce valley-sector band limitation and avoiding spurious spectral leakage between sectors.
We verified convergence of the lowest eigenvalues and $\Delta$ with respect to $N_{\mathrm{BZ}}$ and  $N_{\mathrm{FBZ}}$.

%\bibliography{MK_literature}

\begin{thebibliography}{49}%
\makeatletter
\providecommand \@ifxundefined [1]{%
 \@ifx{#1\undefined}
}%
\providecommand \@ifnum [1]{%
 \ifnum #1\expandafter \@firstoftwo
 \else \expandafter \@secondoftwo
 \fi
}%
\providecommand \@ifx [1]{%
 \ifx #1\expandafter \@firstoftwo
 \else \expandafter \@secondoftwo
 \fi
}%
\providecommand \natexlab [1]{#1}%
\providecommand \enquote  [1]{``#1''}%
\providecommand \bibnamefont  [1]{#1}%
\providecommand \bibfnamefont [1]{#1}%
\providecommand \citenamefont [1]{#1}%
\providecommand \href@noop [0]{\@secondoftwo}%
\providecommand \href [0]{\begingroup \@sanitize@url \@href}%
\providecommand \@href[1]{\@@startlink{#1}\@@href}%
\providecommand \@@href[1]{\endgroup#1\@@endlink}%
\providecommand \@sanitize@url [0]{\catcode `\\12\catcode `\$12\catcode
  `\&12\catcode `\#12\catcode `\^12\catcode `\_12\catcode `\%12\relax}%
\providecommand \@@startlink[1]{}%
\providecommand \@@endlink[0]{}%
\providecommand \url  [0]{\begingroup\@sanitize@url \@url }%
\providecommand \@url [1]{\endgroup\@href {#1}{\urlprefix }}%
\providecommand \urlprefix  [0]{URL }%
\providecommand \Eprint [0]{\href }%
\providecommand \doibase [0]{https://doi.org/}%
\providecommand \selectlanguage [0]{\@gobble}%
\providecommand \bibinfo  [0]{\@secondoftwo}%
\providecommand \bibfield  [0]{\@secondoftwo}%
\providecommand \translation [1]{[#1]}%
\providecommand \BibitemOpen [0]{}%
\providecommand \bibitemStop [0]{}%
\providecommand \bibitemNoStop [0]{.\EOS\space}%
\providecommand \EOS [0]{\spacefactor3000\relax}%
\providecommand \BibitemShut  [1]{\csname bibitem#1\endcsname}%
\let\auto@bib@innerbib\@empty
%</preamble>
\bibitem [{\citenamefont {Burkard}\ \emph {et~al.}(2023)\citenamefont
  {Burkard}, \citenamefont {Ladd}, \citenamefont {Pan}, \citenamefont
  {Nichol},\ and\ \citenamefont {Petta}}]{Burkard2023}%
  \BibitemOpen
  \bibfield  {author} {\bibinfo {author} {\bibfnamefont {G.}~\bibnamefont
  {Burkard}}, \bibinfo {author} {\bibfnamefont {T.~D.}\ \bibnamefont {Ladd}},
  \bibinfo {author} {\bibfnamefont {A.}~\bibnamefont {Pan}}, \bibinfo {author}
  {\bibfnamefont {J.~M.}\ \bibnamefont {Nichol}},\ and\ \bibinfo {author}
  {\bibfnamefont {J.~R.}\ \bibnamefont {Petta}},\ }\bibfield  {title} {\bibinfo
  {title} {Semiconductor spin qubits},\ }\href
  {https://doi.org/10.1103/RevModPhys.95.025003} {\bibfield  {journal}
  {\bibinfo  {journal} {Rev. Mod. Phys.}\ }\textbf {\bibinfo {volume} {95}},\
  \bibinfo {pages} {025003} (\bibinfo {year} {2023})}\BibitemShut {NoStop}%
\bibitem [{\citenamefont {Zwanenburg}\ \emph {et~al.}(2013)\citenamefont
  {Zwanenburg}, \citenamefont {Dzurak}, \citenamefont {Morello}, \citenamefont
  {Simmons}, \citenamefont {Hollenberg}, \citenamefont {Klimeck}, \citenamefont
  {Rogge}, \citenamefont {Coppersmith},\ and\ \citenamefont
  {Eriksson}}]{Zwanenburg2013}%
  \BibitemOpen
  \bibfield  {author} {\bibinfo {author} {\bibfnamefont {F.~A.}\ \bibnamefont
  {Zwanenburg}}, \bibinfo {author} {\bibfnamefont {A.~S.}\ \bibnamefont
  {Dzurak}}, \bibinfo {author} {\bibfnamefont {A.}~\bibnamefont {Morello}},
  \bibinfo {author} {\bibfnamefont {M.~Y.}\ \bibnamefont {Simmons}}, \bibinfo
  {author} {\bibfnamefont {L.~C.~L.}\ \bibnamefont {Hollenberg}}, \bibinfo
  {author} {\bibfnamefont {G.}~\bibnamefont {Klimeck}}, \bibinfo {author}
  {\bibfnamefont {S.}~\bibnamefont {Rogge}}, \bibinfo {author} {\bibfnamefont
  {S.~N.}\ \bibnamefont {Coppersmith}},\ and\ \bibinfo {author} {\bibfnamefont
  {M.~A.}\ \bibnamefont {Eriksson}},\ }\bibfield  {title} {\bibinfo {title}
  {Silicon quantum electronics},\ }\href
  {https://doi.org/10.1103/RevModPhys.85.961} {\bibfield  {journal} {\bibinfo
  {journal} {Rev. Mod. Phys.}\ }\textbf {\bibinfo {volume} {85}},\ \bibinfo
  {pages} {961} (\bibinfo {year} {2013})}\BibitemShut {NoStop}%
\bibitem [{\citenamefont {Yoneda}\ \emph {et~al.}(2017)\citenamefont {Yoneda},
  \citenamefont {Takeda}, \citenamefont {Otsuka}, \citenamefont {Nakajima},
  \citenamefont {Delbecq}, \citenamefont {Allison}, \citenamefont {Honda},
  \citenamefont {Kodera}, \citenamefont {Oda}, \citenamefont {Hoshi},
  \citenamefont {Usami}, \citenamefont {Itoh},\ and\ \citenamefont
  {Tarucha}}]{Yoneda2017}%
  \BibitemOpen
  \bibfield  {author} {\bibinfo {author} {\bibfnamefont {J.}~\bibnamefont
  {Yoneda}}, \bibinfo {author} {\bibfnamefont {K.}~\bibnamefont {Takeda}},
  \bibinfo {author} {\bibfnamefont {T.}~\bibnamefont {Otsuka}}, \bibinfo
  {author} {\bibfnamefont {T.}~\bibnamefont {Nakajima}}, \bibinfo {author}
  {\bibfnamefont {M.~R.}\ \bibnamefont {Delbecq}}, \bibinfo {author}
  {\bibfnamefont {G.}~\bibnamefont {Allison}}, \bibinfo {author} {\bibfnamefont
  {T.}~\bibnamefont {Honda}}, \bibinfo {author} {\bibfnamefont
  {T.}~\bibnamefont {Kodera}}, \bibinfo {author} {\bibfnamefont
  {S.}~\bibnamefont {Oda}}, \bibinfo {author} {\bibfnamefont {Y.}~\bibnamefont
  {Hoshi}}, \bibinfo {author} {\bibfnamefont {N.}~\bibnamefont {Usami}},
  \bibinfo {author} {\bibfnamefont {K.~M.}\ \bibnamefont {Itoh}},\ and\
  \bibinfo {author} {\bibfnamefont {S.}~\bibnamefont {Tarucha}},\ }\bibfield
  {title} {\bibinfo {title} {A quantum-dot spin qubit with coherence limited by
  charge noise and fidelity higher than 99.9\%},\ }\href
  {https://doi.org/10.1038/s41565-017-0014-x} {\bibfield  {journal} {\bibinfo
  {journal} {Nat. Nanotechnol.}\ }\textbf {\bibinfo {volume} {13}},\ \bibinfo
  {pages} {102} (\bibinfo {year} {2017})}\BibitemShut {NoStop}%
\bibitem [{\citenamefont {Tyryshkin}\ \emph {et~al.}(2011)\citenamefont
  {Tyryshkin}, \citenamefont {Tojo}, \citenamefont {Morton}, \citenamefont
  {Riemann}, \citenamefont {Abrosimov}, \citenamefont {Becker}, \citenamefont
  {Pohl}, \citenamefont {Schenkel}, \citenamefont {Thewalt}, \citenamefont
  {Itoh},\ and\ \citenamefont {Lyon}}]{Tyryshkin2011}%
  \BibitemOpen
  \bibfield  {author} {\bibinfo {author} {\bibfnamefont {A.~M.}\ \bibnamefont
  {Tyryshkin}}, \bibinfo {author} {\bibfnamefont {S.}~\bibnamefont {Tojo}},
  \bibinfo {author} {\bibfnamefont {J.~J.~L.}\ \bibnamefont {Morton}}, \bibinfo
  {author} {\bibfnamefont {H.}~\bibnamefont {Riemann}}, \bibinfo {author}
  {\bibfnamefont {N.~V.}\ \bibnamefont {Abrosimov}}, \bibinfo {author}
  {\bibfnamefont {P.}~\bibnamefont {Becker}}, \bibinfo {author} {\bibfnamefont
  {H.-J.}\ \bibnamefont {Pohl}}, \bibinfo {author} {\bibfnamefont
  {T.}~\bibnamefont {Schenkel}}, \bibinfo {author} {\bibfnamefont {M.~L.~W.}\
  \bibnamefont {Thewalt}}, \bibinfo {author} {\bibfnamefont {K.~M.}\
  \bibnamefont {Itoh}},\ and\ \bibinfo {author} {\bibfnamefont {S.~A.}\
  \bibnamefont {Lyon}},\ }\bibfield  {title} {\bibinfo {title} {Electron spin
  coherence exceeding seconds in high-purity silicon},\ }\href
  {https://doi.org/10.1038/nmat3182} {\bibfield  {journal} {\bibinfo  {journal}
  {Nat. Mater.}\ }\textbf {\bibinfo {volume} {11}},\ \bibinfo {pages} {143}
  (\bibinfo {year} {2011})}\BibitemShut {NoStop}%
\bibitem [{\citenamefont {Koch}\ \emph {et~al.}(2025)\citenamefont {Koch},
  \citenamefont {Godfrin}, \citenamefont {Adam}, \citenamefont {Ferrero},
  \citenamefont {Schroller}, \citenamefont {Glaeser}, \citenamefont {Kubicek},
  \citenamefont {Li}, \citenamefont {Loo}, \citenamefont {Massar},
  \citenamefont {Simion}, \citenamefont {Wan}, \citenamefont {De~Greve},\ and\
  \citenamefont {Wernsdorfer}}]{Koch2025}%
  \BibitemOpen
  \bibfield  {author} {\bibinfo {author} {\bibfnamefont {T.}~\bibnamefont
  {Koch}}, \bibinfo {author} {\bibfnamefont {C.}~\bibnamefont {Godfrin}},
  \bibinfo {author} {\bibfnamefont {V.}~\bibnamefont {Adam}}, \bibinfo {author}
  {\bibfnamefont {J.}~\bibnamefont {Ferrero}}, \bibinfo {author} {\bibfnamefont
  {D.}~\bibnamefont {Schroller}}, \bibinfo {author} {\bibfnamefont
  {N.}~\bibnamefont {Glaeser}}, \bibinfo {author} {\bibfnamefont
  {S.}~\bibnamefont {Kubicek}}, \bibinfo {author} {\bibfnamefont
  {R.}~\bibnamefont {Li}}, \bibinfo {author} {\bibfnamefont {R.}~\bibnamefont
  {Loo}}, \bibinfo {author} {\bibfnamefont {S.}~\bibnamefont {Massar}},
  \bibinfo {author} {\bibfnamefont {G.}~\bibnamefont {Simion}}, \bibinfo
  {author} {\bibfnamefont {D.}~\bibnamefont {Wan}}, \bibinfo {author}
  {\bibfnamefont {K.}~\bibnamefont {De~Greve}},\ and\ \bibinfo {author}
  {\bibfnamefont {W.}~\bibnamefont {Wernsdorfer}},\ }\bibfield  {title}
  {\bibinfo {title} {Industrial 300 mm wafer processed spin qubits in natural
  silicon/silicon-germanium},\ }\href
  {https://doi.org/10.1038/s41534-025-01016-x} {\bibfield  {journal} {\bibinfo
  {journal} {npj Quantum Inf.}\ }\textbf {\bibinfo {volume} {11}},\ \bibinfo
  {pages} {59} (\bibinfo {year} {2025})}\BibitemShut {NoStop}%
\bibitem [{\citenamefont {Neyens}\ \emph {et~al.}(2024)\citenamefont {Neyens},
  \citenamefont {Zietz}, \citenamefont {Watson}, \citenamefont {Luthi},
  \citenamefont {Nethwewala}, \citenamefont {George}, \citenamefont {Henry},
  \citenamefont {Islam}, \citenamefont {Wagner}, \citenamefont {Borjans},
  \citenamefont {Connors}, \citenamefont {Corrigan}, \citenamefont {Curry},
  \citenamefont {Keith}, \citenamefont {Kotlyar}, \citenamefont {Lampert},
  \citenamefont {M\k{a}dzik}, \citenamefont {Millard}, \citenamefont
  {Mohiyaddin}, \citenamefont {Pellerano}, \citenamefont {Pillarisetty},
  \citenamefont {Ramsey}, \citenamefont {Savytskyy}, \citenamefont {Schaal},
  \citenamefont {Zheng}, \citenamefont {Ziegler}, \citenamefont {Bishop},
  \citenamefont {Bojarski}, \citenamefont {Roberts},\ and\ \citenamefont
  {Clarke}}]{Neyens2024}%
  \BibitemOpen
  \bibfield  {author} {\bibinfo {author} {\bibfnamefont {S.}~\bibnamefont
  {Neyens}}, \bibinfo {author} {\bibfnamefont {O.~K.}\ \bibnamefont {Zietz}},
  \bibinfo {author} {\bibfnamefont {T.~F.}\ \bibnamefont {Watson}}, \bibinfo
  {author} {\bibfnamefont {F.}~\bibnamefont {Luthi}}, \bibinfo {author}
  {\bibfnamefont {A.}~\bibnamefont {Nethwewala}}, \bibinfo {author}
  {\bibfnamefont {H.~C.}\ \bibnamefont {George}}, \bibinfo {author}
  {\bibfnamefont {E.}~\bibnamefont {Henry}}, \bibinfo {author} {\bibfnamefont
  {M.}~\bibnamefont {Islam}}, \bibinfo {author} {\bibfnamefont {A.~J.}\
  \bibnamefont {Wagner}}, \bibinfo {author} {\bibfnamefont {F.}~\bibnamefont
  {Borjans}}, \bibinfo {author} {\bibfnamefont {E.~J.}\ \bibnamefont
  {Connors}}, \bibinfo {author} {\bibfnamefont {J.}~\bibnamefont {Corrigan}},
  \bibinfo {author} {\bibfnamefont {M.~J.}\ \bibnamefont {Curry}}, \bibinfo
  {author} {\bibfnamefont {D.}~\bibnamefont {Keith}}, \bibinfo {author}
  {\bibfnamefont {R.}~\bibnamefont {Kotlyar}}, \bibinfo {author} {\bibfnamefont
  {L.~F.}\ \bibnamefont {Lampert}}, \bibinfo {author} {\bibfnamefont {M.~T.}\
  \bibnamefont {M\k{a}dzik}}, \bibinfo {author} {\bibfnamefont
  {K.}~\bibnamefont {Millard}}, \bibinfo {author} {\bibfnamefont {F.~A.}\
  \bibnamefont {Mohiyaddin}}, \bibinfo {author} {\bibfnamefont
  {S.}~\bibnamefont {Pellerano}}, \bibinfo {author} {\bibfnamefont
  {R.}~\bibnamefont {Pillarisetty}}, \bibinfo {author} {\bibfnamefont
  {M.}~\bibnamefont {Ramsey}}, \bibinfo {author} {\bibfnamefont
  {R.}~\bibnamefont {Savytskyy}}, \bibinfo {author} {\bibfnamefont
  {S.}~\bibnamefont {Schaal}}, \bibinfo {author} {\bibfnamefont
  {G.}~\bibnamefont {Zheng}}, \bibinfo {author} {\bibfnamefont
  {J.}~\bibnamefont {Ziegler}}, \bibinfo {author} {\bibfnamefont {N.~C.}\
  \bibnamefont {Bishop}}, \bibinfo {author} {\bibfnamefont {S.}~\bibnamefont
  {Bojarski}}, \bibinfo {author} {\bibfnamefont {J.}~\bibnamefont {Roberts}},\
  and\ \bibinfo {author} {\bibfnamefont {J.~S.}\ \bibnamefont {Clarke}},\
  }\bibfield  {title} {\bibinfo {title} {Probing single electrons across 300-mm
  spin qubit wafers},\ }\href {https://doi.org/10.1038/s41586-024-07275-6}
  {\bibfield  {journal} {\bibinfo  {journal} {Nature}\ }\textbf {\bibinfo
  {volume} {629}},\ \bibinfo {pages} {80} (\bibinfo {year} {2024})}\BibitemShut
  {NoStop}%
\bibitem [{\citenamefont {George}\ \emph {et~al.}(2025)\citenamefont {George},
  \citenamefont {M\k{a}dzik}, \citenamefont {Henry}, \citenamefont {Wagner},
  \citenamefont {Islam}, \citenamefont {Borjans}, \citenamefont {Connors},
  \citenamefont {Corrigan}, \citenamefont {Curry}, \citenamefont {Harper},
  \citenamefont {Keith}, \citenamefont {Lampert}, \citenamefont {Luthi},
  \citenamefont {Mohiyaddin}, \citenamefont {Murcia}, \citenamefont {Nair},
  \citenamefont {Nahm}, \citenamefont {Nethwewala}, \citenamefont {Neyens},
  \citenamefont {Raharjo}, \citenamefont {Rogan}, \citenamefont {Savytskyy},
  \citenamefont {Watson}, \citenamefont {Ziegler}, \citenamefont {Zietz},
  \citenamefont {Pillarisetty}, \citenamefont {Bishop}, \citenamefont
  {Bojarski}, \citenamefont {Roberts},\ and\ \citenamefont
  {Clarke}}]{George2025}%
  \BibitemOpen
  \bibfield  {author} {\bibinfo {author} {\bibfnamefont {H.~C.}\ \bibnamefont
  {George}}, \bibinfo {author} {\bibfnamefont {M.~T.}\ \bibnamefont
  {M\k{a}dzik}}, \bibinfo {author} {\bibfnamefont {E.~M.}\ \bibnamefont
  {Henry}}, \bibinfo {author} {\bibfnamefont {A.~J.}\ \bibnamefont {Wagner}},
  \bibinfo {author} {\bibfnamefont {M.~M.}\ \bibnamefont {Islam}}, \bibinfo
  {author} {\bibfnamefont {F.}~\bibnamefont {Borjans}}, \bibinfo {author}
  {\bibfnamefont {E.~J.}\ \bibnamefont {Connors}}, \bibinfo {author}
  {\bibfnamefont {J.}~\bibnamefont {Corrigan}}, \bibinfo {author}
  {\bibfnamefont {M.}~\bibnamefont {Curry}}, \bibinfo {author} {\bibfnamefont
  {M.~K.}\ \bibnamefont {Harper}}, \bibinfo {author} {\bibfnamefont
  {D.}~\bibnamefont {Keith}}, \bibinfo {author} {\bibfnamefont
  {L.}~\bibnamefont {Lampert}}, \bibinfo {author} {\bibfnamefont
  {F.}~\bibnamefont {Luthi}}, \bibinfo {author} {\bibfnamefont {F.~A.}\
  \bibnamefont {Mohiyaddin}}, \bibinfo {author} {\bibfnamefont
  {S.}~\bibnamefont {Murcia}}, \bibinfo {author} {\bibfnamefont
  {R.}~\bibnamefont {Nair}}, \bibinfo {author} {\bibfnamefont {R.}~\bibnamefont
  {Nahm}}, \bibinfo {author} {\bibfnamefont {A.}~\bibnamefont {Nethwewala}},
  \bibinfo {author} {\bibfnamefont {S.}~\bibnamefont {Neyens}}, \bibinfo
  {author} {\bibfnamefont {R.~D.}\ \bibnamefont {Raharjo}}, \bibinfo {author}
  {\bibfnamefont {C.}~\bibnamefont {Rogan}}, \bibinfo {author} {\bibfnamefont
  {R.}~\bibnamefont {Savytskyy}}, \bibinfo {author} {\bibfnamefont {T.~F.}\
  \bibnamefont {Watson}}, \bibinfo {author} {\bibfnamefont {J.}~\bibnamefont
  {Ziegler}}, \bibinfo {author} {\bibfnamefont {O.~K.}\ \bibnamefont {Zietz}},
  \bibinfo {author} {\bibfnamefont {R.}~\bibnamefont {Pillarisetty}}, \bibinfo
  {author} {\bibfnamefont {N.~C.}\ \bibnamefont {Bishop}}, \bibinfo {author}
  {\bibfnamefont {S.~A.}\ \bibnamefont {Bojarski}}, \bibinfo {author}
  {\bibfnamefont {J.}~\bibnamefont {Roberts}},\ and\ \bibinfo {author}
  {\bibfnamefont {J.~S.}\ \bibnamefont {Clarke}},\ }\bibfield  {title}
  {\bibinfo {title} {12-spin-qubit arrays fabricated on a 300 mm semiconductor
  manufacturing line},\ }\href {https://doi.org/10.1021/acs.nanolett.4c05205}
  {\bibfield  {journal} {\bibinfo  {journal} {Nano Letters}\ }\textbf {\bibinfo
  {volume} {25}},\ \bibinfo {pages} {793} (\bibinfo {year} {2025})}\BibitemShut
  {NoStop}%
\bibitem [{\citenamefont {Mills}\ \emph {et~al.}(2022)\citenamefont {Mills},
  \citenamefont {Guinn}, \citenamefont {Gullans}, \citenamefont {Sigillito},
  \citenamefont {Feldman}, \citenamefont {Nielsen},\ and\ \citenamefont
  {Petta}}]{Mills2022}%
  \BibitemOpen
  \bibfield  {author} {\bibinfo {author} {\bibfnamefont {A.~R.}\ \bibnamefont
  {Mills}}, \bibinfo {author} {\bibfnamefont {C.~R.}\ \bibnamefont {Guinn}},
  \bibinfo {author} {\bibfnamefont {M.~J.}\ \bibnamefont {Gullans}}, \bibinfo
  {author} {\bibfnamefont {A.~J.}\ \bibnamefont {Sigillito}}, \bibinfo {author}
  {\bibfnamefont {M.~M.}\ \bibnamefont {Feldman}}, \bibinfo {author}
  {\bibfnamefont {E.}~\bibnamefont {Nielsen}},\ and\ \bibinfo {author}
  {\bibfnamefont {J.~R.}\ \bibnamefont {Petta}},\ }\bibfield  {title} {\bibinfo
  {title} {Two-qubit silicon quantum processor with operation fidelity
  exceeding 99\%},\ }\href {https://doi.org/10.1126/sciadv.abn5130} {\bibfield
  {journal} {\bibinfo  {journal} {Sci. Adv.}\ }\textbf {\bibinfo {volume}
  {8}},\ \bibinfo {pages} {eabn5130} (\bibinfo {year} {2022})}\BibitemShut
  {NoStop}%
\bibitem [{\citenamefont {Noiri}\ \emph {et~al.}(2022)\citenamefont {Noiri},
  \citenamefont {Takeda}, \citenamefont {Nakajima}, \citenamefont {Kobayashi},
  \citenamefont {Sammak}, \citenamefont {Scappucci},\ and\ \citenamefont
  {Tarucha}}]{Noiri2022}%
  \BibitemOpen
  \bibfield  {author} {\bibinfo {author} {\bibfnamefont {A.}~\bibnamefont
  {Noiri}}, \bibinfo {author} {\bibfnamefont {K.}~\bibnamefont {Takeda}},
  \bibinfo {author} {\bibfnamefont {T.}~\bibnamefont {Nakajima}}, \bibinfo
  {author} {\bibfnamefont {T.}~\bibnamefont {Kobayashi}}, \bibinfo {author}
  {\bibfnamefont {A.}~\bibnamefont {Sammak}}, \bibinfo {author} {\bibfnamefont
  {G.}~\bibnamefont {Scappucci}},\ and\ \bibinfo {author} {\bibfnamefont
  {S.}~\bibnamefont {Tarucha}},\ }\bibfield  {title} {\bibinfo {title} {Fast
  universal quantum gate above the fault-tolerance threshold in silicon},\
  }\href {https://doi.org/10.1038/s41586-021-04182-y} {\bibfield  {journal}
  {\bibinfo  {journal} {Nature}\ }\textbf {\bibinfo {volume} {601}},\ \bibinfo
  {pages} {338} (\bibinfo {year} {2022})}\BibitemShut {NoStop}%
\bibitem [{\citenamefont {Xue}\ \emph {et~al.}(2022)\citenamefont {Xue},
  \citenamefont {Russ}, \citenamefont {Samkharadze}, \citenamefont {Undseth},
  \citenamefont {Sammak}, \citenamefont {Scappucci},\ and\ \citenamefont
  {Vandersypen}}]{Xue2022}%
  \BibitemOpen
  \bibfield  {author} {\bibinfo {author} {\bibfnamefont {X.}~\bibnamefont
  {Xue}}, \bibinfo {author} {\bibfnamefont {M.}~\bibnamefont {Russ}}, \bibinfo
  {author} {\bibfnamefont {N.}~\bibnamefont {Samkharadze}}, \bibinfo {author}
  {\bibfnamefont {B.}~\bibnamefont {Undseth}}, \bibinfo {author} {\bibfnamefont
  {A.}~\bibnamefont {Sammak}}, \bibinfo {author} {\bibfnamefont
  {G.}~\bibnamefont {Scappucci}},\ and\ \bibinfo {author} {\bibfnamefont
  {L.~M.~K.}\ \bibnamefont {Vandersypen}},\ }\bibfield  {title} {\bibinfo
  {title} {Quantum logic with spin qubits crossing the surface code
  threshold},\ }\href {https://doi.org/10.1038/s41586-021-04273-w} {\bibfield
  {journal} {\bibinfo  {journal} {Nature}\ }\textbf {\bibinfo {volume} {601}},\
  \bibinfo {pages} {343} (\bibinfo {year} {2022})}\BibitemShut {NoStop}%
\bibitem [{\citenamefont {Yang}\ \emph {et~al.}(2013)\citenamefont {Yang},
  \citenamefont {Rossi}, \citenamefont {Ruskov}, \citenamefont {Lai},
  \citenamefont {Mohiyaddin}, \citenamefont {Lee}, \citenamefont {Tahan},
  \citenamefont {Klimeck}, \citenamefont {Morello},\ and\ \citenamefont
  {Dzurak}}]{Yang2013}%
  \BibitemOpen
  \bibfield  {author} {\bibinfo {author} {\bibfnamefont {C.~H.}\ \bibnamefont
  {Yang}}, \bibinfo {author} {\bibfnamefont {A.}~\bibnamefont {Rossi}},
  \bibinfo {author} {\bibfnamefont {R.}~\bibnamefont {Ruskov}}, \bibinfo
  {author} {\bibfnamefont {N.~S.}\ \bibnamefont {Lai}}, \bibinfo {author}
  {\bibfnamefont {F.~A.}\ \bibnamefont {Mohiyaddin}}, \bibinfo {author}
  {\bibfnamefont {S.}~\bibnamefont {Lee}}, \bibinfo {author} {\bibfnamefont
  {C.}~\bibnamefont {Tahan}}, \bibinfo {author} {\bibfnamefont
  {G.}~\bibnamefont {Klimeck}}, \bibinfo {author} {\bibfnamefont
  {A.}~\bibnamefont {Morello}},\ and\ \bibinfo {author} {\bibfnamefont {A.~S.}\
  \bibnamefont {Dzurak}},\ }\bibfield  {title} {\bibinfo {title} {Spin-valley
  lifetimes in a silicon quantum dot with tunable valley splitting},\ }\href
  {https://doi.org/10.1038/ncomms3069} {\bibfield  {journal} {\bibinfo
  {journal} {Nat. Commun.}\ }\textbf {\bibinfo {volume} {4}},\ \bibinfo {pages}
  {2069} (\bibinfo {year} {2013})}\BibitemShut {NoStop}%
\bibitem [{\citenamefont {Zhang}\ \emph {et~al.}(2020)\citenamefont {Zhang},
  \citenamefont {Hu}, \citenamefont {Li}, \citenamefont {Jing}, \citenamefont
  {Zhou}, \citenamefont {Ma}, \citenamefont {Ni}, \citenamefont {Luo},
  \citenamefont {Cao}, \citenamefont {Wang}, \citenamefont {Hu}, \citenamefont
  {Jiang}, \citenamefont {Guo},\ and\ \citenamefont {Guo}}]{Zhang2020}%
  \BibitemOpen
  \bibfield  {author} {\bibinfo {author} {\bibfnamefont {X.}~\bibnamefont
  {Zhang}}, \bibinfo {author} {\bibfnamefont {R.-Z.}\ \bibnamefont {Hu}},
  \bibinfo {author} {\bibfnamefont {H.-O.}\ \bibnamefont {Li}}, \bibinfo
  {author} {\bibfnamefont {F.-M.}\ \bibnamefont {Jing}}, \bibinfo {author}
  {\bibfnamefont {Y.}~\bibnamefont {Zhou}}, \bibinfo {author} {\bibfnamefont
  {R.-L.}\ \bibnamefont {Ma}}, \bibinfo {author} {\bibfnamefont
  {M.}~\bibnamefont {Ni}}, \bibinfo {author} {\bibfnamefont {G.}~\bibnamefont
  {Luo}}, \bibinfo {author} {\bibfnamefont {G.}~\bibnamefont {Cao}}, \bibinfo
  {author} {\bibfnamefont {G.-L.}\ \bibnamefont {Wang}}, \bibinfo {author}
  {\bibfnamefont {X.}~\bibnamefont {Hu}}, \bibinfo {author} {\bibfnamefont
  {H.-W.}\ \bibnamefont {Jiang}}, \bibinfo {author} {\bibfnamefont {G.-C.}\
  \bibnamefont {Guo}},\ and\ \bibinfo {author} {\bibfnamefont {G.-P.}\
  \bibnamefont {Guo}},\ }\bibfield  {title} {\bibinfo {title} {Giant anisotropy
  of spin relaxation and spin-valley mixing in a silicon quantum dot},\ }\href
  {https://doi.org/10.1103/PhysRevLett.124.257701} {\bibfield  {journal}
  {\bibinfo  {journal} {Phys. Rev. Lett.}\ }\textbf {\bibinfo {volume} {124}},\
  \bibinfo {pages} {257701} (\bibinfo {year} {2020})}\BibitemShut {NoStop}%
\bibitem [{\citenamefont {Losert}\ \emph {et~al.}(2024)\citenamefont {Losert},
  \citenamefont {Oberl\"ander}, \citenamefont {Teske}, \citenamefont {Volmer},
  \citenamefont {Schreiber}, \citenamefont {Bluhm}, \citenamefont
  {Coppersmith},\ and\ \citenamefont {Friesen}}]{Losert2024}%
  \BibitemOpen
  \bibfield  {author} {\bibinfo {author} {\bibfnamefont {M.~P.}\ \bibnamefont
  {Losert}}, \bibinfo {author} {\bibfnamefont {M.}~\bibnamefont
  {Oberl\"ander}}, \bibinfo {author} {\bibfnamefont {J.~D.}\ \bibnamefont
  {Teske}}, \bibinfo {author} {\bibfnamefont {M.}~\bibnamefont {Volmer}},
  \bibinfo {author} {\bibfnamefont {L.~R.}\ \bibnamefont {Schreiber}}, \bibinfo
  {author} {\bibfnamefont {H.}~\bibnamefont {Bluhm}}, \bibinfo {author}
  {\bibfnamefont {S.}~\bibnamefont {Coppersmith}},\ and\ \bibinfo {author}
  {\bibfnamefont {M.}~\bibnamefont {Friesen}},\ }\bibfield  {title} {\bibinfo
  {title} {Strategies for enhancing spin-shuttling fidelities in {Si}/{SiGe}
  quantum wells with random-alloy disorder},\ }\href
  {https://doi.org/10.1103/PRXQuantum.5.040322} {\bibfield  {journal} {\bibinfo
   {journal} {PRX Quantum}\ }\textbf {\bibinfo {volume} {5}},\ \bibinfo {pages}
  {040322} (\bibinfo {year} {2024})}\BibitemShut {NoStop}%
\bibitem [{\citenamefont {Degli~Esposti}\ \emph {et~al.}(2024)\citenamefont
  {Degli~Esposti}, \citenamefont {Stehouwer}, \citenamefont {G\"{u}l},
  \citenamefont {Samkharadze}, \citenamefont {D\'{e}prez}, \citenamefont
  {Meyer}, \citenamefont {Meijer}, \citenamefont {Tryputen}, \citenamefont
  {Karwal}, \citenamefont {Botifoll}, \citenamefont {Arbiol}, \citenamefont
  {Amitonov}, \citenamefont {Vandersypen}, \citenamefont {Sammak},
  \citenamefont {Veldhorst},\ and\ \citenamefont
  {Scappucci}}]{DegliEsposti2024}%
  \BibitemOpen
  \bibfield  {author} {\bibinfo {author} {\bibfnamefont {D.}~\bibnamefont
  {Degli~Esposti}}, \bibinfo {author} {\bibfnamefont {L.~E.~A.}\ \bibnamefont
  {Stehouwer}}, \bibinfo {author} {\bibfnamefont {O.}~\bibnamefont {G\"{u}l}},
  \bibinfo {author} {\bibfnamefont {N.}~\bibnamefont {Samkharadze}}, \bibinfo
  {author} {\bibfnamefont {C.}~\bibnamefont {D\'{e}prez}}, \bibinfo {author}
  {\bibfnamefont {M.}~\bibnamefont {Meyer}}, \bibinfo {author} {\bibfnamefont
  {I.~N.}\ \bibnamefont {Meijer}}, \bibinfo {author} {\bibfnamefont
  {L.}~\bibnamefont {Tryputen}}, \bibinfo {author} {\bibfnamefont
  {S.}~\bibnamefont {Karwal}}, \bibinfo {author} {\bibfnamefont
  {M.}~\bibnamefont {Botifoll}}, \bibinfo {author} {\bibfnamefont
  {J.}~\bibnamefont {Arbiol}}, \bibinfo {author} {\bibfnamefont {S.~V.}\
  \bibnamefont {Amitonov}}, \bibinfo {author} {\bibfnamefont {L.~M.~K.}\
  \bibnamefont {Vandersypen}}, \bibinfo {author} {\bibfnamefont
  {A.}~\bibnamefont {Sammak}}, \bibinfo {author} {\bibfnamefont
  {M.}~\bibnamefont {Veldhorst}},\ and\ \bibinfo {author} {\bibfnamefont
  {G.}~\bibnamefont {Scappucci}},\ }\bibfield  {title} {\bibinfo {title} {Low
  disorder and high valley splitting in silicon},\ }\href
  {https://doi.org/10.1038/s41534-024-00826-9} {\bibfield  {journal} {\bibinfo
  {journal} {npj Quantum Inf.}\ }\textbf {\bibinfo {volume} {10}},\ \bibinfo
  {pages} {32} (\bibinfo {year} {2024})}\BibitemShut {NoStop}%
\bibitem [{\citenamefont {McJunkin}\ \emph {et~al.}(2021)\citenamefont
  {McJunkin}, \citenamefont {MacQuarrie}, \citenamefont {Tom}, \citenamefont
  {Neyens}, \citenamefont {Dodson}, \citenamefont {Thorgrimsson}, \citenamefont
  {Corrigan}, \citenamefont {Ercan}, \citenamefont {Savage}, \citenamefont
  {Lagally}, \citenamefont {Joynt}, \citenamefont {Coppersmith}, \citenamefont
  {Friesen},\ and\ \citenamefont {Eriksson}}]{McJunkin2021}%
  \BibitemOpen
  \bibfield  {author} {\bibinfo {author} {\bibfnamefont {T.}~\bibnamefont
  {McJunkin}}, \bibinfo {author} {\bibfnamefont {E.~R.}\ \bibnamefont
  {MacQuarrie}}, \bibinfo {author} {\bibfnamefont {L.}~\bibnamefont {Tom}},
  \bibinfo {author} {\bibfnamefont {S.~F.}\ \bibnamefont {Neyens}}, \bibinfo
  {author} {\bibfnamefont {J.~P.}\ \bibnamefont {Dodson}}, \bibinfo {author}
  {\bibfnamefont {B.}~\bibnamefont {Thorgrimsson}}, \bibinfo {author}
  {\bibfnamefont {J.}~\bibnamefont {Corrigan}}, \bibinfo {author}
  {\bibfnamefont {H.~E.}\ \bibnamefont {Ercan}}, \bibinfo {author}
  {\bibfnamefont {D.~E.}\ \bibnamefont {Savage}}, \bibinfo {author}
  {\bibfnamefont {M.~G.}\ \bibnamefont {Lagally}}, \bibinfo {author}
  {\bibfnamefont {R.}~\bibnamefont {Joynt}}, \bibinfo {author} {\bibfnamefont
  {S.~N.}\ \bibnamefont {Coppersmith}}, \bibinfo {author} {\bibfnamefont
  {M.}~\bibnamefont {Friesen}},\ and\ \bibinfo {author} {\bibfnamefont {M.~A.}\
  \bibnamefont {Eriksson}},\ }\bibfield  {title} {\bibinfo {title} {Valley
  splittings in {Si}/{SiGe} quantum dots with a germanium spike in the silicon
  well},\ }\href {https://doi.org/10.1103/PhysRevB.104.085406} {\bibfield
  {journal} {\bibinfo  {journal} {Phys. Rev. B}\ }\textbf {\bibinfo {volume}
  {104}},\ \bibinfo {pages} {085406} (\bibinfo {year} {2021})}\BibitemShut
  {NoStop}%
\bibitem [{\citenamefont {McJunkin}\ \emph {et~al.}(2022)\citenamefont
  {McJunkin}, \citenamefont {Harpt}, \citenamefont {Feng}, \citenamefont
  {Losert}, \citenamefont {Rahman}, \citenamefont {Dodson}, \citenamefont
  {Wolfe}, \citenamefont {Savage}, \citenamefont {Lagally}, \citenamefont
  {Coppersmith}, \citenamefont {Friesen}, \citenamefont {Joynt},\ and\
  \citenamefont {Eriksson}}]{McJunkin2022}%
  \BibitemOpen
  \bibfield  {author} {\bibinfo {author} {\bibfnamefont {T.}~\bibnamefont
  {McJunkin}}, \bibinfo {author} {\bibfnamefont {B.}~\bibnamefont {Harpt}},
  \bibinfo {author} {\bibfnamefont {Y.}~\bibnamefont {Feng}}, \bibinfo {author}
  {\bibfnamefont {M.~P.}\ \bibnamefont {Losert}}, \bibinfo {author}
  {\bibfnamefont {R.}~\bibnamefont {Rahman}}, \bibinfo {author} {\bibfnamefont
  {J.~P.}\ \bibnamefont {Dodson}}, \bibinfo {author} {\bibfnamefont {M.~A.}\
  \bibnamefont {Wolfe}}, \bibinfo {author} {\bibfnamefont {D.~E.}\ \bibnamefont
  {Savage}}, \bibinfo {author} {\bibfnamefont {M.~G.}\ \bibnamefont {Lagally}},
  \bibinfo {author} {\bibfnamefont {S.~N.}\ \bibnamefont {Coppersmith}},
  \bibinfo {author} {\bibfnamefont {M.}~\bibnamefont {Friesen}}, \bibinfo
  {author} {\bibfnamefont {R.}~\bibnamefont {Joynt}},\ and\ \bibinfo {author}
  {\bibfnamefont {M.~A.}\ \bibnamefont {Eriksson}},\ }\bibfield  {title}
  {\bibinfo {title} {{SiGe} quantum wells with oscillating {Ge} concentrations
  for quantum dot qubits},\ }\href {https://doi.org/10.1038/s41467-022-35510-z}
  {\bibfield  {journal} {\bibinfo  {journal} {Nat. Commun.}\ }\textbf {\bibinfo
  {volume} {13}},\ \bibinfo {pages} {7777} (\bibinfo {year}
  {2022})}\BibitemShut {NoStop}%
\bibitem [{\citenamefont {Feng}\ and\ \citenamefont {Joynt}(2022)}]{Feng2022}%
  \BibitemOpen
  \bibfield  {author} {\bibinfo {author} {\bibfnamefont {Y.}~\bibnamefont
  {Feng}}\ and\ \bibinfo {author} {\bibfnamefont {R.}~\bibnamefont {Joynt}},\
  }\bibfield  {title} {\bibinfo {title} {Enhanced valley splitting in {Si}
  layers with oscillatory {Ge} concentration},\ }\href
  {https://doi.org/10.1103/PhysRevB.106.085304} {\bibfield  {journal} {\bibinfo
   {journal} {Phys. Rev. B}\ }\textbf {\bibinfo {volume} {106}},\ \bibinfo
  {pages} {085304} (\bibinfo {year} {2022})}\BibitemShut {NoStop}%
\bibitem [{\citenamefont {Woods}\ \emph {et~al.}(2024)\citenamefont {Woods},
  \citenamefont {Soomro}, \citenamefont {Joseph}, \citenamefont {Frink},
  \citenamefont {Joynt}, \citenamefont {Eriksson},\ and\ \citenamefont
  {Friesen}}]{Woods2024}%
  \BibitemOpen
  \bibfield  {author} {\bibinfo {author} {\bibfnamefont {B.~D.}\ \bibnamefont
  {Woods}}, \bibinfo {author} {\bibfnamefont {H.}~\bibnamefont {Soomro}},
  \bibinfo {author} {\bibfnamefont {E.~S.}\ \bibnamefont {Joseph}}, \bibinfo
  {author} {\bibfnamefont {C.~C.~D.}\ \bibnamefont {Frink}}, \bibinfo {author}
  {\bibfnamefont {R.}~\bibnamefont {Joynt}}, \bibinfo {author} {\bibfnamefont
  {M.~A.}\ \bibnamefont {Eriksson}},\ and\ \bibinfo {author} {\bibfnamefont
  {M.}~\bibnamefont {Friesen}},\ }\bibfield  {title} {\bibinfo {title}
  {Coupling conduction-band valleys in modulated {SiGe} heterostructures via
  shear strain},\ }\href {https://doi.org/10.1038/s41534-024-00853-6}
  {\bibfield  {journal} {\bibinfo  {journal} {npj Quantum Inf.}\ }\textbf
  {\bibinfo {volume} {10}},\ \bibinfo {pages} {54} (\bibinfo {year}
  {2024})}\BibitemShut {NoStop}%
\bibitem [{\citenamefont {Gradwohl}\ \emph {et~al.}(2025)\citenamefont
  {Gradwohl}, \citenamefont {Cvitkovich}, \citenamefont {Lu}, \citenamefont
  {Koelling}, \citenamefont {Oezkent}, \citenamefont {Liu}, \citenamefont
  {Waldh\"{o}r}, \citenamefont {Grasser}, \citenamefont {Niquet}, \citenamefont
  {Albrecht}, \citenamefont {Richter}, \citenamefont {Moutanabbir},\ and\
  \citenamefont {Martin}}]{Gradwohl2025}%
  \BibitemOpen
  \bibfield  {author} {\bibinfo {author} {\bibfnamefont {K.-P.}\ \bibnamefont
  {Gradwohl}}, \bibinfo {author} {\bibfnamefont {L.}~\bibnamefont
  {Cvitkovich}}, \bibinfo {author} {\bibfnamefont {C.-H.}\ \bibnamefont {Lu}},
  \bibinfo {author} {\bibfnamefont {S.}~\bibnamefont {Koelling}}, \bibinfo
  {author} {\bibfnamefont {M.}~\bibnamefont {Oezkent}}, \bibinfo {author}
  {\bibfnamefont {Y.}~\bibnamefont {Liu}}, \bibinfo {author} {\bibfnamefont
  {D.}~\bibnamefont {Waldh\"{o}r}}, \bibinfo {author} {\bibfnamefont
  {T.}~\bibnamefont {Grasser}}, \bibinfo {author} {\bibfnamefont {Y.-M.}\
  \bibnamefont {Niquet}}, \bibinfo {author} {\bibfnamefont {M.}~\bibnamefont
  {Albrecht}}, \bibinfo {author} {\bibfnamefont {C.}~\bibnamefont {Richter}},
  \bibinfo {author} {\bibfnamefont {O.}~\bibnamefont {Moutanabbir}},\ and\
  \bibinfo {author} {\bibfnamefont {J.}~\bibnamefont {Martin}},\ }\bibfield
  {title} {\bibinfo {title} {Enhanced nanoscale {Ge} concentration oscillations
  in {Si}/{SiGe} quantum well through controlled segregation},\ }\href
  {https://doi.org/10.1021/acs.nanolett.4c05326} {\bibfield  {journal}
  {\bibinfo  {journal} {Nano Lett.}\ }\textbf {\bibinfo {volume} {25}},\
  \bibinfo {pages} {4204} (\bibinfo {year} {2025})}\BibitemShut {NoStop}%
\bibitem [{\citenamefont {Niquet}\ \emph {et~al.}(2009)\citenamefont {Niquet},
  \citenamefont {Rideau}, \citenamefont {Tavernier}, \citenamefont {Jaouen},\
  and\ \citenamefont {Blase}}]{Niquet2009}%
  \BibitemOpen
  \bibfield  {author} {\bibinfo {author} {\bibfnamefont {Y.~M.}\ \bibnamefont
  {Niquet}}, \bibinfo {author} {\bibfnamefont {D.}~\bibnamefont {Rideau}},
  \bibinfo {author} {\bibfnamefont {C.}~\bibnamefont {Tavernier}}, \bibinfo
  {author} {\bibfnamefont {H.}~\bibnamefont {Jaouen}},\ and\ \bibinfo {author}
  {\bibfnamefont {X.}~\bibnamefont {Blase}},\ }\bibfield  {title} {\bibinfo
  {title} {Onsite matrix elements of the tight-binding hamiltonian of a
  strained crystal: Application to silicon, germanium, and their alloys},\
  }\href {https://doi.org/10.1103/PhysRevB.79.245201} {\bibfield  {journal}
  {\bibinfo  {journal} {Phys. Rev. B}\ }\textbf {\bibinfo {volume} {79}},\
  \bibinfo {pages} {245201} (\bibinfo {year} {2009})}\BibitemShut {NoStop}%
\bibitem [{\citenamefont {Losert}\ \emph {et~al.}(2023)\citenamefont {Losert},
  \citenamefont {Eriksson}, \citenamefont {Joynt}, \citenamefont {Rahman},
  \citenamefont {Scappucci}, \citenamefont {Coppersmith},\ and\ \citenamefont
  {Friesen}}]{Losert2023}%
  \BibitemOpen
  \bibfield  {author} {\bibinfo {author} {\bibfnamefont {M.~P.}\ \bibnamefont
  {Losert}}, \bibinfo {author} {\bibfnamefont {M.~A.}\ \bibnamefont
  {Eriksson}}, \bibinfo {author} {\bibfnamefont {R.}~\bibnamefont {Joynt}},
  \bibinfo {author} {\bibfnamefont {R.}~\bibnamefont {Rahman}}, \bibinfo
  {author} {\bibfnamefont {G.}~\bibnamefont {Scappucci}}, \bibinfo {author}
  {\bibfnamefont {S.~N.}\ \bibnamefont {Coppersmith}},\ and\ \bibinfo {author}
  {\bibfnamefont {M.}~\bibnamefont {Friesen}},\ }\bibfield  {title} {\bibinfo
  {title} {Practical strategies for enhancing the valley splitting in
  {Si}/{SiGe} quantum wells},\ }\href
  {https://doi.org/10.1103/PhysRevB.108.125405} {\bibfield  {journal} {\bibinfo
   {journal} {Phys. Rev. B}\ }\textbf {\bibinfo {volume} {108}},\ \bibinfo
  {pages} {125405} (\bibinfo {year} {2023})}\BibitemShut {NoStop}%
\bibitem [{\citenamefont {Cvitkovich}(2024)}]{Cvitkovich2024}%
  \BibitemOpen
  \bibfield  {author} {\bibinfo {author} {\bibfnamefont {L.}~\bibnamefont
  {Cvitkovich}},\ }\emph {\bibinfo {title} {Atomistic Modeling of {Si} Spin
  Qubits From First Principles}},\ \href
  {https://doi.org/10.34726/hss.2024.123305} {\bibinfo {type} {phdthesis}},\
  \bibinfo  {school} {TU Vienna} (\bibinfo {year} {2024})\BibitemShut {NoStop}%
\bibitem [{\citenamefont {Cvitkovich}\ \emph {et~al.}(2026)\citenamefont
  {Cvitkovich}, \citenamefont {Salamone}, \citenamefont {Wilhelmer},
  \citenamefont {Martinez}, \citenamefont {Grasser},\ and\ \citenamefont
  {Niquet}}]{Cvitkovich2026}%
  \BibitemOpen
  \bibfield  {author} {\bibinfo {author} {\bibfnamefont {L.}~\bibnamefont
  {Cvitkovich}}, \bibinfo {author} {\bibfnamefont {T.}~\bibnamefont
  {Salamone}}, \bibinfo {author} {\bibfnamefont {C.}~\bibnamefont {Wilhelmer}},
  \bibinfo {author} {\bibfnamefont {B.}~\bibnamefont {Martinez}}, \bibinfo
  {author} {\bibfnamefont {T.}~\bibnamefont {Grasser}},\ and\ \bibinfo {author}
  {\bibfnamefont {Y.-M.}\ \bibnamefont {Niquet}},\ }\bibfield  {title}
  {\bibinfo {title} {Valley splittings in {Si}/{SiGe} heterostructures from
  first principles},\ }\href {https://doi.org/10.1103/frx3-41bz} {\bibfield
  {journal} {\bibinfo  {journal} {Phys. Rev. B}\ }\textbf {\bibinfo {volume}
  {113}},\ \bibinfo {pages} {035307} (\bibinfo {year} {2026})}\BibitemShut
  {NoStop}%
\bibitem [{\citenamefont {Kohn}\ and\ \citenamefont
  {Luttinger}(1955)}]{Kohn1955}%
  \BibitemOpen
  \bibfield  {author} {\bibinfo {author} {\bibfnamefont {W.}~\bibnamefont
  {Kohn}}\ and\ \bibinfo {author} {\bibfnamefont {J.~M.}\ \bibnamefont
  {Luttinger}},\ }\bibfield  {title} {\bibinfo {title} {Theory of donor states
  in silicon},\ }\href {https://doi.org/https://doi.org/10.1103/PhysRev.98.915}
  {\bibfield  {journal} {\bibinfo  {journal} {Phys. Rev.}\ }\textbf {\bibinfo
  {volume} {98}},\ \bibinfo {pages} {915} (\bibinfo {year} {1955})}\BibitemShut
  {NoStop}%
\bibitem [{\citenamefont {Fritzsche}(1962)}]{Fritzsche1962}%
  \BibitemOpen
  \bibfield  {author} {\bibinfo {author} {\bibfnamefont {H.}~\bibnamefont
  {Fritzsche}},\ }\bibfield  {title} {\bibinfo {title} {Effect of stress on the
  donor wave functions in germanium},\ }\href
  {https://doi.org/10.1103/PhysRev.125.1560} {\bibfield  {journal} {\bibinfo
  {journal} {Phys. Rev.}\ }\textbf {\bibinfo {volume} {125}},\ \bibinfo {pages}
  {1560} (\bibinfo {year} {1962})}\BibitemShut {NoStop}%
\bibitem [{\citenamefont {Baldereschi}(1970)}]{Baldereschi1970}%
  \BibitemOpen
  \bibfield  {author} {\bibinfo {author} {\bibfnamefont {A.}~\bibnamefont
  {Baldereschi}},\ }\bibfield  {title} {\bibinfo {title} {Valley-orbit
  interaction in semiconductors},\ }\href
  {https://doi.org/10.1103/PhysRevB.1.4673} {\bibfield  {journal} {\bibinfo
  {journal} {Phys. Rev. B}\ }\textbf {\bibinfo {volume} {1}},\ \bibinfo {pages}
  {4673} (\bibinfo {year} {1970})}\BibitemShut {NoStop}%
\bibitem [{\citenamefont {Ning}\ and\ \citenamefont {Sah}(1971)}]{Ning1971}%
  \BibitemOpen
  \bibfield  {author} {\bibinfo {author} {\bibfnamefont {T.~H.}\ \bibnamefont
  {Ning}}\ and\ \bibinfo {author} {\bibfnamefont {C.~T.}\ \bibnamefont {Sah}},\
  }\bibfield  {title} {\bibinfo {title} {Multivalley effective-mass
  approximation for donor states in silicon. {I.} {Shallow}-level group-{V}
  impurities},\ }\href {https://doi.org/10.1103/PhysRevB.4.3468} {\bibfield
  {journal} {\bibinfo  {journal} {Phys. Rev. B}\ }\textbf {\bibinfo {volume}
  {4}},\ \bibinfo {pages} {3468} (\bibinfo {year} {1971})}\BibitemShut
  {NoStop}%
\bibitem [{\citenamefont {Pantelides}\ and\ \citenamefont
  {Sah}(1974)}]{Pantelides1974}%
  \BibitemOpen
  \bibfield  {author} {\bibinfo {author} {\bibfnamefont {S.~T.}\ \bibnamefont
  {Pantelides}}\ and\ \bibinfo {author} {\bibfnamefont {C.~T.}\ \bibnamefont
  {Sah}},\ }\bibfield  {title} {\bibinfo {title} {Theory of localized states in
  semiconductors. i. new results using an old method},\ }\href
  {https://doi.org/10.1103/PhysRevB.10.621} {\bibfield  {journal} {\bibinfo
  {journal} {Phys. Rev. B}\ }\textbf {\bibinfo {volume} {10}},\ \bibinfo
  {pages} {621} (\bibinfo {year} {1974})}\BibitemShut {NoStop}%
\bibitem [{\citenamefont {Shindo}\ and\ \citenamefont
  {Nara}(1976)}]{Shindo1976}%
  \BibitemOpen
  \bibfield  {author} {\bibinfo {author} {\bibfnamefont {K.}~\bibnamefont
  {Shindo}}\ and\ \bibinfo {author} {\bibfnamefont {H.}~\bibnamefont {Nara}},\
  }\bibfield  {title} {\bibinfo {title} {The effective mass equation for the
  multi-valley semiconductors},\ }\href {https://doi.org/10.1143/JPSJ.40.1640}
  {\bibfield  {journal} {\bibinfo  {journal} {J. Phys. Soc. Jpn.}\ }\textbf
  {\bibinfo {volume} {40}},\ \bibinfo {pages} {1640} (\bibinfo {year}
  {1976})}\BibitemShut {NoStop}%
\bibitem [{\citenamefont {Debernardi}\ \emph {et~al.}(2006)\citenamefont
  {Debernardi}, \citenamefont {Baldereschi},\ and\ \citenamefont
  {Fanciulli}}]{Debernardi2006}%
  \BibitemOpen
  \bibfield  {author} {\bibinfo {author} {\bibfnamefont {A.}~\bibnamefont
  {Debernardi}}, \bibinfo {author} {\bibfnamefont {A.}~\bibnamefont
  {Baldereschi}},\ and\ \bibinfo {author} {\bibfnamefont {M.}~\bibnamefont
  {Fanciulli}},\ }\bibfield  {title} {\bibinfo {title} {Computation of the
  stark effect in p impurity states in silicon},\ }\href
  {https://doi.org/10.1103/PhysRevB.74.035202} {\bibfield  {journal} {\bibinfo
  {journal} {Phys. Rev. B}\ }\textbf {\bibinfo {volume} {74}},\ \bibinfo
  {pages} {035202} (\bibinfo {year} {2006})}\BibitemShut {NoStop}%
\bibitem [{\citenamefont {Hui}(2013)}]{Hui2013}%
  \BibitemOpen
  \bibfield  {author} {\bibinfo {author} {\bibfnamefont {H.}~\bibnamefont
  {Hui}},\ }\bibfield  {title} {\bibinfo {title} {An improved
  effective-mass-theory equation for phosphorus doped in silicon},\ }\href
  {https://doi.org/10.1016/j.ssc.2012.10.023} {\bibfield  {journal} {\bibinfo
  {journal} {Solid State Commun.}\ }\textbf {\bibinfo {volume} {154}},\
  \bibinfo {pages} {19} (\bibinfo {year} {2013})}\BibitemShut {NoStop}%
\bibitem [{\citenamefont {Pendo}\ \emph {et~al.}(2013)\citenamefont {Pendo},
  \citenamefont {Handberg}, \citenamefont {Smelyanskiy},\ and\ \citenamefont
  {Petukhov}}]{Pendo2013}%
  \BibitemOpen
  \bibfield  {author} {\bibinfo {author} {\bibfnamefont {L.}~\bibnamefont
  {Pendo}}, \bibinfo {author} {\bibfnamefont {E.~M.}\ \bibnamefont {Handberg}},
  \bibinfo {author} {\bibfnamefont {V.~N.}\ \bibnamefont {Smelyanskiy}},\ and\
  \bibinfo {author} {\bibfnamefont {A.~G.}\ \bibnamefont {Petukhov}},\
  }\bibfield  {title} {\bibinfo {title} {Large stark effect for li donor spins
  in si},\ }\href {https://doi.org/10.1103/PhysRevB.88.045307} {\bibfield
  {journal} {\bibinfo  {journal} {Phys. Rev. B}\ }\textbf {\bibinfo {volume}
  {88}},\ \bibinfo {pages} {045307} (\bibinfo {year} {2013})}\BibitemShut
  {NoStop}%
\bibitem [{\citenamefont {Gamble}\ \emph {et~al.}(2015)\citenamefont {Gamble},
  \citenamefont {Jacobson}, \citenamefont {Nielsen}, \citenamefont {Baczewski},
  \citenamefont {Moussa}, \citenamefont {Montaño},\ and\ \citenamefont
  {Muller}}]{Gamble2015}%
  \BibitemOpen
  \bibfield  {author} {\bibinfo {author} {\bibfnamefont {J.~K.}\ \bibnamefont
  {Gamble}}, \bibinfo {author} {\bibfnamefont {N.~T.}\ \bibnamefont
  {Jacobson}}, \bibinfo {author} {\bibfnamefont {E.}~\bibnamefont {Nielsen}},
  \bibinfo {author} {\bibfnamefont {A.~D.}\ \bibnamefont {Baczewski}}, \bibinfo
  {author} {\bibfnamefont {J.~E.}\ \bibnamefont {Moussa}}, \bibinfo {author}
  {\bibfnamefont {I.}~\bibnamefont {Montaño}},\ and\ \bibinfo {author}
  {\bibfnamefont {R.~P.}\ \bibnamefont {Muller}},\ }\bibfield  {title}
  {\bibinfo {title} {Multivalley effective mass theory simulation of donors in
  silicon},\ }\href {https://doi.org/10.1103/PhysRevB.91.235318} {\bibfield
  {journal} {\bibinfo  {journal} {Phys. Rev. B}\ }\textbf {\bibinfo {volume}
  {91}},\ \bibinfo {pages} {235318} (\bibinfo {year} {2015})}\BibitemShut
  {NoStop}%
\bibitem [{\citenamefont {Burt}(1988)}]{Burt1988}%
  \BibitemOpen
  \bibfield  {author} {\bibinfo {author} {\bibfnamefont {M.~G.}\ \bibnamefont
  {Burt}},\ }\bibfield  {title} {\bibinfo {title} {An exact formulation of the
  envelope function method for the determination of electronic states in
  semiconductor microstructures},\ }\href
  {https://doi.org/10.1088/0268-1242/3/8/003} {\bibfield  {journal} {\bibinfo
  {journal} {Semicond. Sci. Tech.}\ }\textbf {\bibinfo {volume} {3}},\ \bibinfo
  {pages} {739} (\bibinfo {year} {1988})}\BibitemShut {NoStop}%
\bibitem [{\citenamefont {Burt}(1994)}]{Burt1994}%
  \BibitemOpen
  \bibfield  {author} {\bibinfo {author} {\bibfnamefont {M.~G.}\ \bibnamefont
  {Burt}},\ }\bibfield  {title} {\bibinfo {title} {Direct derivation of
  effective-mass equations for microstructures with atomically abrupt
  boundaries},\ }\href {https://doi.org/10.1103/PhysRevB.50.7518} {\bibfield
  {journal} {\bibinfo  {journal} {Phys. Rev. B}\ }\textbf {\bibinfo {volume}
  {50}},\ \bibinfo {pages} {7518} (\bibinfo {year} {1994})}\BibitemShut
  {NoStop}%
\bibitem [{\citenamefont {Foreman}(1995)}]{Foreman1995}%
  \BibitemOpen
  \bibfield  {author} {\bibinfo {author} {\bibfnamefont {B.~A.}\ \bibnamefont
  {Foreman}},\ }\bibfield  {title} {\bibinfo {title} {Exact effective-mass
  theory for heterostructures},\ }\href
  {https://doi.org/10.1103/PhysRevB.52.12241} {\bibfield  {journal} {\bibinfo
  {journal} {Phys. Rev. B}\ }\textbf {\bibinfo {volume} {52}},\ \bibinfo
  {pages} {12241} (\bibinfo {year} {1995})}\BibitemShut {NoStop}%
\bibitem [{\citenamefont {Foreman}(1996)}]{Foreman1996}%
  \BibitemOpen
  \bibfield  {author} {\bibinfo {author} {\bibfnamefont {B.~A.}\ \bibnamefont
  {Foreman}},\ }\bibfield  {title} {\bibinfo {title} {Envelope-function
  formalism for electrons in abrupt heterostructures with material-dependent
  basis functions},\ }\href {https://doi.org/10.1103/PhysRevB.54.1909}
  {\bibfield  {journal} {\bibinfo  {journal} {Phys. Rev. B}\ }\textbf {\bibinfo
  {volume} {54}},\ \bibinfo {pages} {1909} (\bibinfo {year}
  {1996})}\BibitemShut {NoStop}%
\bibitem [{\citenamefont {Klymenko}\ and\ \citenamefont
  {Remacle}(2014)}]{Klymenko2014}%
  \BibitemOpen
  \bibfield  {author} {\bibinfo {author} {\bibfnamefont {M.~V.}\ \bibnamefont
  {Klymenko}}\ and\ \bibinfo {author} {\bibfnamefont {F.}~\bibnamefont
  {Remacle}},\ }\bibfield  {title} {\bibinfo {title} {Electronic states and
  wavefunctions of diatomic donor molecular ions in silicon: multi-valley
  envelope function theory},\ }\href
  {https://doi.org/10.1088/0953-8984/26/6/065302} {\bibfield  {journal}
  {\bibinfo  {journal} {J. Phys. Condens. Matter}\ }\textbf {\bibinfo {volume}
  {26}},\ \bibinfo {pages} {065302} (\bibinfo {year} {2014})}\BibitemShut
  {NoStop}%
\bibitem [{\citenamefont {Klymenko}\ \emph {et~al.}(2015)\citenamefont
  {Klymenko}, \citenamefont {Rogge},\ and\ \citenamefont
  {Remacle}}]{Klymenko2015}%
  \BibitemOpen
  \bibfield  {author} {\bibinfo {author} {\bibfnamefont {M.~V.}\ \bibnamefont
  {Klymenko}}, \bibinfo {author} {\bibfnamefont {S.}~\bibnamefont {Rogge}},\
  and\ \bibinfo {author} {\bibfnamefont {F.}~\bibnamefont {Remacle}},\
  }\bibfield  {title} {\bibinfo {title} {Multivalley envelope function
  equations and effective potentials for phosphorus impurity in silicon},\
  }\href {https://doi.org/10.1103/PhysRevB.92.195302} {\bibfield  {journal}
  {\bibinfo  {journal} {Phys. Rev. B}\ }\textbf {\bibinfo {volume} {92}},\
  \bibinfo {pages} {195302} (\bibinfo {year} {2015})}\BibitemShut {NoStop}%
\bibitem [{\citenamefont {Thayil}\ \emph
  {et~al.}(2025{\natexlab{a}})\citenamefont {Thayil}, \citenamefont
  {Ermoneit},\ and\ \citenamefont {Kantner}}]{Thayil2025}%
  \BibitemOpen
  \bibfield  {author} {\bibinfo {author} {\bibfnamefont {A.}~\bibnamefont
  {Thayil}}, \bibinfo {author} {\bibfnamefont {L.}~\bibnamefont {Ermoneit}},\
  and\ \bibinfo {author} {\bibfnamefont {M.}~\bibnamefont {Kantner}},\
  }\bibfield  {title} {\bibinfo {title} {Theory of valley splitting in
  {Si}/{SiGe} spin-qubits: Interplay of strain, resonances and random alloy
  disorder},\ }\href {https://doi.org/10.1103/4sdz-f9cr} {\bibfield  {journal}
  {\bibinfo  {journal} {Phys. Rev. B}\ }\textbf {\bibinfo {volume} {112}},\
  \bibinfo {pages} {115303} (\bibinfo {year} {2025}{\natexlab{a}})}\BibitemShut
  {NoStop}%
\bibitem [{\citenamefont {Van~de Walle}\ and\ \citenamefont
  {Martin}(1986)}]{VandeWalle1986}%
  \BibitemOpen
  \bibfield  {author} {\bibinfo {author} {\bibfnamefont {C.~G.}\ \bibnamefont
  {Van~de Walle}}\ and\ \bibinfo {author} {\bibfnamefont {R.~M.}\ \bibnamefont
  {Martin}},\ }\bibfield  {title} {\bibinfo {title} {Theoretical calculations
  of heterojunction discontinuities in the {Si}/{Ge} system},\ }\href
  {https://doi.org/10.1103/PhysRevB.34.5621} {\bibfield  {journal} {\bibinfo
  {journal} {Phys. Rev. B}\ }\textbf {\bibinfo {volume} {34}},\ \bibinfo
  {pages} {5621} (\bibinfo {year} {1986})}\BibitemShut {NoStop}%
\bibitem [{\citenamefont {Sch\"{a}ffler}(1997)}]{Schaeffler1997}%
  \BibitemOpen
  \bibfield  {author} {\bibinfo {author} {\bibfnamefont {F.}~\bibnamefont
  {Sch\"{a}ffler}},\ }\bibfield  {title} {\bibinfo {title} {High-mobility {Si}
  and {Ge} structures},\ }\href {https://doi.org/10.1088/0268-1242/12/12/001}
  {\bibfield  {journal} {\bibinfo  {journal} {Semicond. Sci. Technol.}\
  }\textbf {\bibinfo {volume} {12}},\ \bibinfo {pages} {1515} (\bibinfo {year}
  {1997})}\BibitemShut {NoStop}%
\bibitem [{\citenamefont {Rieger}\ and\ \citenamefont
  {Vogl}(1993)}]{Rieger1993}%
  \BibitemOpen
  \bibfield  {author} {\bibinfo {author} {\bibfnamefont {M.~M.}\ \bibnamefont
  {Rieger}}\ and\ \bibinfo {author} {\bibfnamefont {P.}~\bibnamefont {Vogl}},\
  }\bibfield  {title} {\bibinfo {title} {Electronic-band parameters in strained
  {Si}$_{1-x}${Ge}$_x$ alloys on {S}i$_{1-y}${Ge}$_y$ substrates},\ }\href
  {https://doi.org/10.1103/PhysRevB.48.14276} {\bibfield  {journal} {\bibinfo
  {journal} {Phys. Rev. B}\ }\textbf {\bibinfo {volume} {48}},\ \bibinfo
  {pages} {14276} (\bibinfo {year} {1993})}\BibitemShut {NoStop}%
\bibitem [{\citenamefont {Ungersboeck}\ \emph {et~al.}(2007)\citenamefont
  {Ungersboeck}, \citenamefont {Dhar}, \citenamefont {Karlowatz}, \citenamefont
  {Sverdlov}, \citenamefont {Kosina},\ and\ \citenamefont
  {Selberherr}}]{Ungersboeck2007b}%
  \BibitemOpen
  \bibfield  {author} {\bibinfo {author} {\bibfnamefont {E.}~\bibnamefont
  {Ungersboeck}}, \bibinfo {author} {\bibfnamefont {S.}~\bibnamefont {Dhar}},
  \bibinfo {author} {\bibfnamefont {G.}~\bibnamefont {Karlowatz}}, \bibinfo
  {author} {\bibfnamefont {V.}~\bibnamefont {Sverdlov}}, \bibinfo {author}
  {\bibfnamefont {H.}~\bibnamefont {Kosina}},\ and\ \bibinfo {author}
  {\bibfnamefont {S.}~\bibnamefont {Selberherr}},\ }\bibfield  {title}
  {\bibinfo {title} {The effect of general strain on the band structure and
  electron mobility of silicon},\ }\href
  {https://doi.org/10.1109/TED.2007.902880} {\bibfield  {journal} {\bibinfo
  {journal} {IEEE Trans. Electron Devices}\ }\textbf {\bibinfo {volume} {54}},\
  \bibinfo {pages} {2183} (\bibinfo {year} {2007})}\BibitemShut {NoStop}%
\bibitem [{\citenamefont {Saraiva}\ \emph {et~al.}(2009)\citenamefont
  {Saraiva}, \citenamefont {Calder\'{o}n}, \citenamefont {Hu}, \citenamefont
  {Das~Sarma},\ and\ \citenamefont {Koiller}}]{Saraiva2009}%
  \BibitemOpen
  \bibfield  {author} {\bibinfo {author} {\bibfnamefont {A.~L.}\ \bibnamefont
  {Saraiva}}, \bibinfo {author} {\bibfnamefont {M.~J.}\ \bibnamefont
  {Calder\'{o}n}}, \bibinfo {author} {\bibfnamefont {X.}~\bibnamefont {Hu}},
  \bibinfo {author} {\bibfnamefont {S.}~\bibnamefont {Das~Sarma}},\ and\
  \bibinfo {author} {\bibfnamefont {B.}~\bibnamefont {Koiller}},\ }\bibfield
  {title} {\bibinfo {title} {Physical mechanisms of interface-mediated
  intervalley coupling in {Si}},\ }\href
  {https://doi.org/10.1103/PhysRevB.80.081305} {\bibfield  {journal} {\bibinfo
  {journal} {Phys. Rev. B}\ }\textbf {\bibinfo {volume} {80}},\ \bibinfo
  {pages} {081305} (\bibinfo {year} {2009})}\BibitemShut {NoStop}%
\bibitem [{\citenamefont {Friesen}\ \emph {et~al.}(2007)\citenamefont
  {Friesen}, \citenamefont {Chutia}, \citenamefont {Tahan},\ and\ \citenamefont
  {Coppersmith}}]{Friesen2007}%
  \BibitemOpen
  \bibfield  {author} {\bibinfo {author} {\bibfnamefont {M.}~\bibnamefont
  {Friesen}}, \bibinfo {author} {\bibfnamefont {S.}~\bibnamefont {Chutia}},
  \bibinfo {author} {\bibfnamefont {C.}~\bibnamefont {Tahan}},\ and\ \bibinfo
  {author} {\bibfnamefont {S.~N.}\ \bibnamefont {Coppersmith}},\ }\bibfield
  {title} {\bibinfo {title} {Valley splitting theory of {SiGe}/{Si}/{SiGe}
  quantum wells},\ }\href {https://doi.org/10.1103/PhysRevB.75.115318}
  {\bibfield  {journal} {\bibinfo  {journal} {Phys. Rev. B}\ }\textbf {\bibinfo
  {volume} {75}},\ \bibinfo {pages} {115318} (\bibinfo {year}
  {2007})}\BibitemShut {NoStop}%
\bibitem [{\citenamefont {Thayil}\ \emph
  {et~al.}(2025{\natexlab{b}})\citenamefont {Thayil}, \citenamefont {Ermoneit},
  \citenamefont {Schreiber}, \citenamefont {Koprucki},\ and\ \citenamefont
  {Kantner}}]{Thayil2025c}%
  \BibitemOpen
  \bibfield  {author} {\bibinfo {author} {\bibfnamefont {A.}~\bibnamefont
  {Thayil}}, \bibinfo {author} {\bibfnamefont {L.}~\bibnamefont {Ermoneit}},
  \bibinfo {author} {\bibfnamefont {L.~R.}\ \bibnamefont {Schreiber}}, \bibinfo
  {author} {\bibfnamefont {T.}~\bibnamefont {Koprucki}},\ and\ \bibinfo
  {author} {\bibfnamefont {M.}~\bibnamefont {Kantner}},\ }\bibfield  {title}
  {\bibinfo {title} {Optimization of {Si}/{SiGe} heterostructures for large and
  robust valley splitting in silicon qubits},\ }\bibfield  {journal} {\bibinfo
  {journal} {arXiv:2512.18064}\ }\href
  {https://doi.org/10.48550/arXiv.2512.18064} {10.48550/arXiv.2512.18064}
  (\bibinfo {year} {2025}{\natexlab{b}})\BibitemShut {NoStop}%
\bibitem [{\citenamefont {Salamone}\ \emph {et~al.}(2025)\citenamefont
  {Salamone}, \citenamefont {Diaz}, \citenamefont {Li}, \citenamefont
  {Cvitkovich},\ and\ \citenamefont {Niquet}}]{Salamone2025}%
  \BibitemOpen
  \bibfield  {author} {\bibinfo {author} {\bibfnamefont {T.}~\bibnamefont
  {Salamone}}, \bibinfo {author} {\bibfnamefont {B.~M.}\ \bibnamefont {Diaz}},
  \bibinfo {author} {\bibfnamefont {J.}~\bibnamefont {Li}}, \bibinfo {author}
  {\bibfnamefont {L.}~\bibnamefont {Cvitkovich}},\ and\ \bibinfo {author}
  {\bibfnamefont {Y.-M.}\ \bibnamefont {Niquet}},\ }\bibfield  {title}
  {\bibinfo {title} {Valley physics in the two bands k.p model for {SiGe}
  heterostructures and spin qubits},\ }\bibfield  {journal} {\bibinfo
  {journal} {arxiv}\ }\href {https://doi.org/10.48550/arXiv.2511.20153}
  {10.48550/arXiv.2511.20153} (\bibinfo {year} {2025})\BibitemShut {NoStop}%
\bibitem [{\citenamefont {Ermoneit}\ \emph {et~al.}()\citenamefont {Ermoneit},
  \citenamefont {Thayil}, \citenamefont {Koprucki},\ and\ \citenamefont
  {Kantner}}]{Ermoneit2026b}%
  \BibitemOpen
  \bibfield  {author} {\bibinfo {author} {\bibfnamefont {L.}~\bibnamefont
  {Ermoneit}}, \bibinfo {author} {\bibfnamefont {A.}~\bibnamefont {Thayil}},
  \bibinfo {author} {\bibfnamefont {T.}~\bibnamefont {Koprucki}},\ and\
  \bibinfo {author} {\bibfnamefont {M.}~\bibnamefont {Kantner}},\ }\href
  {https://github.com/kantner/multi-valley-envelope} {}\bibinfo {note}
  {{GitHub} repository with MATLAB simulation code.
  \url{https://github.com/kantner/multi-valley-envelope}}\BibitemShut {NoStop}%
\end{thebibliography}

\end{document}